\documentclass[twocolumn]{article}

\pdfoutput=1

\usepackage{styling/common}
\usepackage{styling/sizing}

\begin{document}

\title{\bfseries Can 100 Machines Agree?}

\date{}

\author[$\dagger$]{Rachid Guerraoui}
\author[$\dagger$]{Jad Hamza}
\author[$\dagger$]{Dragos-Adrian Seredinschi}
\author[$\ddag$]{Marko Vukolic\thanks{Authors appear in alphabetical order.}}
\affil[$\dagger$]{EPFL}
\affil[$\ddag$]{IBM Research -- Zurich}

\maketitle



\begin{abstract}

Agreement protocols have been typically deployed at small scale, e.g., using three to five machines.
This is because these protocols seem to suffer from a \emph{sharp performance decay}.
More specifically, as the size of a deployment---i.e., degree of replication---increases, the protocol performance greatly decreases.
There is not much experimental evidence for this decay in practice, however, notably for larger system sizes, e.g., beyond a handful of machines.

In this paper we execute agreement protocols on up to $100$ machines and observe on their performance decay.
We consider well-known agreement protocols part of mature systems, such as Apache ZooKeeper, etcd, and BFT-Smart, as well as a chain and a novel ring-based agreement protocol which we implement ourselves.

We provide empirical evidence that current agreement protocols execute gracefully on $100$ machines.
We observe that throughput decay is initially sharp (consistent with previous observations); but intriguingly---as each system grows beyond a few tens of replicas---the decay dampens.
For chain- and ring-based replication, this decay is slower than for the other systems.
The positive takeaway from our evaluation is that mature agreement protocol implementations can sustain \emph{out-of-the-box} $300$ to $500$ requests per second when executing on $100$ replicas on a wide-area public cloud platform.
Chain- and ring-based replication can reach between $4K$ and $11K$ (up to $20x$ improvements) depending on the fault assumptions.
\end{abstract}







\section{Introduction}

An agreement (or consensus) protocol~\cite{dwork88dls,lam98paxos} allows nodes in a distributed system to agree on every action they take, and hence maintain their state consistently.
Agreement protocols are essential for providing strong consistency, or linearizability \cite{HerlihyW90}, in distributed systems, and such protocols are able to withstand even arbitrary (i.e., Byzantine) failures~\cite{castro2002,lamp82byzantine}.

According to distributed systems folklore, agreement protocols are too expensive if deployed on more than a handful of machines~\cite{bezerra2016strong,cowling2006hq,Kermarrec2000,hu10zookeeper}.
Indeed, there is a tradeoff between performance and fault-tolerance, and some performance decay is inherent in strongly-consistent systems~\cite{abd05quorumupdate}.
When growing in size, such a system can tolerate more faults---but performance does not scale accordingly.
This is because the system replicas (more precisely, a quorum of them) must agree on every operation.
Hence, as the number of machines increases, the cost for agreement increases.

A few workarounds exist to deal with this performance decay.
First, some systems ensure strong consistency for only a small, critical subset of their state (e.g., configuration), while the rest of the system has a scalable design under weaker guarantees~\cite{adya2010centrifuge,gh03gfs,qi13espresso}.
The critical part of the system builds on mature agreement protocols such as ZooKeeper~\cite{hu10zookeeper}, etcd~\cite{etcd,ongaro14search}, Consul~\cite{mishra1993consul}, or Boxwood~\cite{maccormick2004boxwood}.

A second workaround is sharding~\cite{glenden11scatter,ab13sharding}.
In this case, the service state is broken down into disjoint shards, each shard running as a separate agreement cluster~\cite{bezerra2016strong,co13spanner}.
Additional mechanism for cross-shard coordination, such as 2PC~\cite{ba11megastore}, ensures that the whole system stays consistent.

Yet a third workaround consists in abandoning strong consistency, eschewing agreement protocols~\cite{bre12cap,Vogels09}.
Avoiding agreement protocols is sometimes possible for certain data types, e.g., CRDTs~\cite{agu11das,du18beat,gup16nonconsensus,sh11crdt}.
But for solving certain problems, notably general state machine replication (SMR), databases, or smart contracts, an agreement building block is necessary~\cite{co13spanner,guerra19cnc,sou18byzantine}.


Briefly, the purpose of these workarounds is to avoid executing agreement at a larger scale.
Consequently, agreement protocols have almost never been deployed, in practice, on more than a few machines, typically three to five~\cite{co13spanner}.
Today there is not much empirical evidence of their throughput decay.
For instance, we do not know how ZooKeeper or PBFT~\cite{castro2002} perform with, say, $100$ machines.
In fact, anecdotal evidence suggests that agreement protocols often \emph{do not work} altogether beyond a few machines~\cite{clement09making,gueta2018sbft,sou18byzantine}.

This question---of performance decay for agreement protocols---is not only of academic interest.
For example, agreement protocols are important in decentralized services: they stand at the heart of distributed ledger applications in permissioned environments~\cite{hyperledger,sou18byzantine}.
For these applications, SMR protocols are expected to run on at least a few \emph{tens} of machines~\cite{Croman16,vuko15quest}.
Another example is in a sharded design: under the Byzantine fault-tolerance model, shards cannot be too small, otherwise an adversary can easily compromise the whole system.
In this case, it is critical for each shard to comprise tens or hundreds of machines~\cite{koko17omniledger,Kermarrec2000}.
But agreement protocols struggle from the ``they do not scale'' stigma and the lack of experiments around their performance decay.




In this chapter we address the void in the literature by deploying and observing how the size of a system (executing agreement) impacts its performance.
We focus on SMR systems, since agreement protocols most commonly appear in such systems.
Our primary goal is to obtain empirical evidence of how SMR performance decays in practice at larger size, and hopefully allay some of the skepticism around the ability of such systems to execute across tens or hundreds of replicas (i.e., machines).
We deploy and evaluate \emph{five} SMR systems on up to $100$ replicas and report on our results.

The first three systems we study are well-known SMR implementations:
ZooKeeper~\cite{hu10zookeeper} and etcd~\cite{etcd}, which are crash fault-tolerant (CFT), and BFT-Smart~\cite{bessani2014state}, which is Byzantine fault-tolerant (BFT).
Consistently with previous observations~\cite{bezerra2016strong,cowling2006hq,hu10zookeeper}, we observe that their throughput decays sharply at small scale.
The interesting part is that this sharp decay does not persist.
Overall their throughput follows an inversely proportional decay rate, and
throughput decay dampens as systems get larger, e.g., beyond ${\sim}40$ replicas.

ZooKeeper, etcd, and BFT-Smart execute most efficiently---obtain best performance---when deployed at their smallest size, i.e., running on $3$ replicas ($4$ for BFT-Smart).
Throughput drops to $50\%$ of its best value at $11$ replicas.
When running on $50$ replicas, the throughput decays to almost $10\%$.
On $100$ replicas the throughput drops to roughly $6\%$ of its best value.
In absolute numbers, these systems sustain $300$ to $500$ \emph{rps} (requests per second) at $100$ replicas on modest hardware in a public cloud platform.
The average latency is below $3.5$s, while the 99th percentile is $6.5$s, even for BFT.

These three systems are hardened SMR implementations and we choose them for their maturity.
We complement the performance observations with a stability study.
Briefly, we seek to understand whether these systems are capable to function despite faults at large scale.
More precisely, we inject a fault in their primary (i.e., leader) replica and evaluate their ability to recover.
We find that ZooKeeper recovers excellently (in a few seconds), indicating
that this system can perform predictably at scale, for instance to implement a replicated system across hundreds of nodes.
The other two systems are slower to recover or have difficulties doing so at $100$ replicas.

The fourth system we investigate is \chainr , based on chain replication~\cite{van04chain}.
This system is throughput-optimized, so it helps us delineate the ideal case, namely, a throughput upper bound.
When growing from $3$ to $100$ replicas, throughput in \chainr decays very slowly, from $15$k \emph{rps} to $11$k \emph{rps} (i.e., to $73\%$ of its best value).
If we place replicas carefully on the wide-area network so as to minimize chain traversal time, \chainr exhibits below $3$ seconds latency.

It can be misleading, however, to praise chain replication as the ideal agreement protocol.
\chainr does not suffer from performance decay as severely as others, indeed---but only in graceful executions (i.e., failure-free and synchronous network).
In non-graceful runs, throughput drops to zero.
This protocol sacrifices availability, because it must reconfigure (pausing execution) to remove any faulty replica~\cite{van04chain}.
This system relies on a synchronous model with fail-stop faults, a strong assumption.
Worse still, the chain is as efficient as its weakest link, so a single straggler can drag down the performance of the whole system.

The fifth system employs a ring overlay (a chain generalization).
We call this system \name , and we design it ourselves.
In contrast to \chainr, this system does not pause execution for reconfiguration, maintaining availability despite asynchrony or faults.

Unlike prior solutions, \name does not rely on reconfiguration~\cite{van04chain,van12byzantine} nor a classic broadcast mode~\cite{aublin15next700,knezevic12high} for masking faults or asynchrony.
Doing so would incur downtime and hurt performance.
Instead, we take the following approach:
Each replica keeps fallback (i.e., redundant) connections to other replicas.
When faulty replicas prevent (or slow down) progress, a correct replica can activate its fallback path(s) to restore progress and maintain availability.
The goal of this simple mechanism in \name is to selectively bypass faults or stragglers on the ring topology (preserving good throughput).

To the best of our knowledge, \name is the first ring-based system to preserve its topology (and availability) despite active faults.
In a $106$-nodes system, \name sustains $6k$ ops/sec when there are no faults (throughput decay to $55\%$ of its best value).
If $F=21$ replicas manifest malicious behavior, then throughput reaches $4k$ ops/sec ($48\%$ decay).
Since \name and \chainr are research prototypes, we do not evaluate their stability, which we leave for future work.

To summarize, in this paper we investigate how performance decays in agreement protocols when increasing their size.
We deploy five SMR systems using at least $100$ replicas in a geo-replicated network.
We observe that, indeed, throughput decays in these systems as a function of system size, but this decay dampens.
Our experiments with chain- and ring-replication show that there are ways to alleviate throughput decay in SMR, informing future designs.

In the rest of this paper, we provide some background, including the SMR systems under our study (\Cref{sec:background}), and then discuss the methodology of our empirical evaluation (\Cref{sec:methodology}).
Next, we present the results of our evaluation on the performance decay of five SMR systems (\Cref{sec:evaluation}).
We take a rather unconventional approach of presenting first the evaluation of \name (in \Cref{sec:evaluation}), and then present its design (\Cref{sec:common-case}).
We also discuss related work (\Cref{sec:discussion-related}), and then conclude (\Cref{sec:conclusions}).

\section{Background \& Motivation}
\label{sec:background}

Consensus protocols often employ a layered design.
In Paxos terminology, for instance, there is a distinction between proposers, acceptors, and learners (or observers)~\cite{van15vive}.
Proposers and acceptors effectively handle the agreement protocol, while learners handle the execution of operations (i.e., client requests).

We are interested in the \emph{agreement} protocol.
This is the one typically encountering scalability issues.
As mentioned earlier, agreement should ideally execute on tens or hundreds of replicas in certain applications, e.g., for decentralized services, or to ensure that shards are resilient against a Byzantine adversary~\cite{Croman16,koko17omniledger,Kermarrec2000,vuko15quest}.

When increasing the size of a system executing agreement, some performance degradation is unavoidable.
This is inherent to replicated systems that ensure strong consistency, because a higher degree of replication (i.e., fault-tolerance) entails a bigger overhead to agree on each request.
But how does performance decay---in a linear manner?
Or does the decay worsen or does it lessen when system size increases?
Both throughput and latency are vital measures of performance, and it is well known that these two are at odds with each other in SMR systems~\cite{guerraoui16icg,lamp03lowerbounds}.
In this paper we focus on throughput, but our study also covers latency results.

We focus on systems with \emph{deterministic} guarantees.
There is a growing body of work proposing agreement protocols for very large systems.
Most of this work, however, offers probabilistic guarantees, e.g., systems designed for cryptocurrencies~\cite{eyal16bitcoinng,gilad2017algorand,koko17omniledger}, or group communication~\cite{Kermarrec2000}.
These probabilistic solutions often employ a committee-based design, where a certain subset of nodes executes agreement on behalf of the whole system.
The protocol which this subset of nodes typically execute is, in fact, a deterministic agreement algorithm.
This can be seen most clearly, for instance, in distributed ledger systems such as ByzCoin~\cite{kogi16byzcoin} or Bitcoin-NG~\cite{eyal16bitcoinng}, where the chosen committee runs a PBFT-like algorithm~\cite{castro2002}.

\begin{table*}[t]
  \caption{Overview of the SMR systems in our study. $^{\ast}$With the tentative execution optimization~\cite{castro2002,sousa15wheat}. $^{\dagger}$This is the worst-case delay, but depending on which replica receives the client request, the message delay can be as low as $N+2$ (as we explain later~\Cref{sec:common-case}).}
  \label{tab:systems}
  \small
  \begin{tabular}{lccccc}
    \toprule
    & ZooKeeper (ZAB~\cite{junq11zab}) & etcd (Raft~\cite{ongaro2014consensus}) & BFT-Smart~\cite{bessani2014state} & \chainr~\cite{van04chain} & \name (\Cref{sec:common-case}) \\
    \midrule
    Synchrony assumptions& partial sync.& partial sync. & partial sync. & sync. & partial sync.\\
    Fault model; system size N & crash; $2F{+}1$ & crash; $2F{+}1$ & Byzantine; $3F{+}1$ & crash; $F{+}1$ & Byzantine; $5F{+}1$\\
    Communication pattern (overlay) & leader-centric &  leader-centric  & leader-centric & chain & ring\\
    Msg. processed by bottleneck node& ${\sim}3N$ & ${\sim}3N$ & ${\sim}4N$ & $1$ & $N+1$\\
    One-way message delays & $4$ & $4$ & $4^{\ast}$ & $N+2$ & $2N+2^{\dagger}$\\
    \bottomrule
  \end{tabular}
\end{table*}

\subsection{SMR Systems in Our Study}
\label{sec:systems}


Our study covers five representative SMR protocols.
We give an overview of these system in~\Cref{tab:systems}, highlighting some of their essential differences.
As can be seen, these systems cover two types of synchrony models (partially synchronous, and synchronous), two fault models (crash and Byzantine), and there are three classes of communication patterns (leader-centric, chain, and ring topology).

In terms of message complexity at the bottleneck node, in the first three protocols the leader does two or three rounds of broadcast.
In \chainr and \name the load is equally distributed across all replicas:
in \chainr each replica sends exactly one message per request, whereas each replica in \name processes $N{+}1$ messages per request.
Finally, the message delays row include communication to and from client.
To sum-up, these five systems cover a wide range of design choices.
We now discuss each system in more detail.


\paragraph{ZooKeeper.} This system is based on ZAB, an atomic broadcast algorithm~\cite{junq11zab}, and is implemented in Java.
We study ZooKeeper rather than ZAB directly, as ZAB is tightly integrated inside ZooKeeper.

\paragraph{etcd.} This system is implemented in Go and is based on the Raft consensus algorithm~\cite{ongaro14search}.
ZAB and Raft share many design points~\cite{ongaro2014consensus}.
A central feature in both is the existence of a leader replica which guides the agreement protocol through a series of broadcast-based rounds.
As our experiments will show, these two systems experience very similar throughput decay (\Cref{sec:evaluation}).

ZooKeeper and etcd are widely used in production and are actively maintained.
They have found adoption both in cluster and multi-datacenter (WAN) environments~\cite{etcduse,an2015wide,bec11leader,fournierzkacross}.
CFT SMR protocols are also relevant to implementing decentralized trust, e.g., in a blockchain.
For instance, a version of Hyperledger Fabric uses Apache Kafka as a CFT consensus algorithm~\cite{kafkaHL}.
These two CFT protocols are interesting in their own right, not necessarily in blockchains, and can also indicate how SMR performs in certain variants of the Byzantine fault model (e.g., XFT model~\cite{liu15xft}).

\paragraph{BFT-Smart.} The third system we study, implemented in Java, provides BFT guarantees~\cite{bessani2014state}.
BFT-Smart is actively used and has been maintained by a team of developers for over eight years, being a default choice for prototyping research algorithms in several groups~\cite{bftshome,liu15xft,Visigoth}.
BFT-Smart is patterned after the seminal PBFT consensus algorithm of Castro and Liskov~\cite{castro2002}.

These three systems described so far employ a \emph{leader-centric} design \cite{bi12spaxos,ongaro2014consensus}, i.e., they rely on the leader replica to carry most of the burden in the agreement protocol.
Specifically, the leader does not only establish a total-order across operations, but also disseminates (via broadcast) those operations to all replicas.
This design simplifies the SMR algorithm~\cite{ongaro2014consensus}.
The disadvantage is that the leader replica (its CPU or bandwidth) is typically the bottleneck~\cite{junq11zab}.

We choose these three systems for their maturity.
These are production-ready (ZooKeeper and etcd) or seasoned implementations (BFT-Smart).
We also study the \emph{stability} of these three systems, i.e., executions where the leader replica crashes, in addition to their performance.
Prototypes, like the next two systems we consider, may deliver better performance, but often do so without vital production-relevant features which can hamper performance.

The fourth and fifth SMR systems we wrote ourselves: \chainr, our prototype of chain replication~\cite{van04chain}, and \name, a new ring-based replication protocol with BFT guarantees.
Both are written in Go.
Chain replication, and in particular its ring-based variants, are provably throughput-optimal in uniform networks~\cite{guerraoui10ring,jalili17ring}.
\footnote{As we discuss later (\Cref{sec:evaluation} and~\Cref{sec:discussion-related}), the WAN testbed we use is not uniform, yet these systems show very good performance.}
In contrast to leader-centric protocols, these systems avoid the bottleneck at the leader because they balance the burden of disseminating operations across multiple system replicas.

\paragraph{\chainr .} We use this system as a baseline, to obtain an ideal upper bound---and what other SMR systems could aim for---both in terms of absolute throughput and throughput decay.
We faithfully implement the common-case with pipelining and batching, to optimize performance~\cite{ongaro2014consensus,santos2012tuning}.

\chainr works in a fail-stop model, i.e., assumes synchrony to be able to mask crash faults~\cite{van04chain}, unlike the other four systems we study in this paper.
Solutions to make chain- or ring-based systems fault-tolerant include an external reconfiguration module, or a special recovery mode~\cite{aublin15next700,ab13sharding,knezevic12high,van12byzantine}.
In such solutions, even simple crashes put the system in a degraded mode, possibly for extended periods of time.


\paragraph{\name .} This system represents our effort in designing a BFT protocol optimized for throughput, which can withstand sub-optimal conditions (e.g., faults, asynchrony) and hence offer improved availability.
We briefly describe \name below, and additionally dedicate a section for full details (\Cref{sec:common-case}).

\name is a ring-based agreement protocol assuming partial synchrony.
When a fault occurs, we mask this by temporarily increasing the fanout at a particular replica.
This is in contrast to prior ring or chain designs, which resort to reconfiguration or recovery.

By default, every replica in \name has a fanout of $1$, i.e., it forwards everything it receives to its immediate successor on ring.
In the worst case, $F$ consecutive replicas on the ring can be faulty.
In this case, the predecessor of all these nodes (a correct replica) increases its fanout to $F+1$, bypassing all $F$ faults.
This way, the successor of these faulty nodes still receives all updates propagating on the ring, and progress is not interrupted.
The system throughput deteriorates when this happens, but not as badly as that of broadcast-based solutions, where one of the replicas---the leader---has a fanout of $2F+1$ (or $3F+1$ for BFT)~\cite{castro2002,hu10zookeeper}.




\section{Methodology}
\label{sec:methodology}

We now discuss the testbed for our study (\Cref{sec:testbeds}), details of the write-only workload (\Cref{sec:workload}), as well as the workload suite we use to conduct experiments (\Cref{sec:workload-suite}).

\subsection{Testbeds}
\label{sec:testbeds}

We consider as testbed the SoftLayer public cloud  platform spanning multiple datacenters~\cite{softlayerDCs}.
We use virtual machines (VMs) equipped with $2$ (virtual) CPU cores and $4$GB RAM.
We use low spec-ed VMs to gain insight in SMR scalability on commodity hardware.
Each client and replica executes in a separate VM.
This separation avoids unnecessary noise in our results, which would happen if client and/or replica processes were contending for local resources.




\paragraph{Network.}
The bandwidth available between different nodes, either clients or replicas, is not a bottleneck: every VM has $100$Mbps bandwidth.
Latencies in SoftLayer range from under $10$ms to almost $200$ms, depending on distance; we consider nine regions of North, Central, and South America.
We use \texttt{ping} to measure the inter-regional latencies (which are symmetric), and present our results in~\Cref{tab:softlayer-latencies}.

\begin{table}[t]
\footnotesize
\begin{tabularx}{\columnwidth}{X | X X X X X X X X}
              & MON & TOR & DAL & SEA & SJC & HOU & MEX & SAO \\
\hline
WDC & 15 & 22 & 31 & 56 & 60 & 39 & 56 & 115 \\
MON &  & 9 & 38 & 61 & 64 & 43 & 63 & 123 \\
TOR &  &  & 30 & 53 & 56 & 37 & 55 & 124 \\
DAL &  &  &  & 40 & 36 & 8 & 25 & 143 \\
SEA &  &  &  &  & 18 & 48 & 65 & 174 \\
SJC &  &  &  &  &  & 44 & 56 & 195 \\
HOU &  &  &  &  &  &  & 30 & 136 \\
MEX &  &  &  &  &  &  &  & 167 \\
\end{tabularx}
\caption{Inter-regional RTT (msec) in SoftLayer. The nine regions are: Washington (WDC),  Montreal (MON), Toronto (TOR), Dallas (DAL), Seattle (SEA), San Jose (SJC), Houston (HOU), Mexico (MEX), and Sao Paolo (SAO).}
\vspace{-.2cm}
\label{tab:softlayer-latencies}
\end{table}


\paragraph{Node placement.}
As~\Cref{tab:softlayer-latencies} illustrates, there is a large disparity in cross-regional latencies.
Consequently, replica and client placement across regions can impact performance.
By default, we always place all clients in Washington.
Spreading clients randomly has no benefit, and would introduce unnecessary variability in the results.
For ZooKeeper, etcd, and BFT-Smart, we place replicas randomly across the nine regions.


Replica placement is particularly important for chain- or ring-based systems.
For instance, in \chainr, client requests propagate from one replica to another, starting from the head of the chain until it reaches the tail; the tail responds to clients.
If we distribute replicas randomly, then requests will pass back and forth between regions, accumulating latencies on the order of seconds or worse.
Random replica placement would be unreasonable.
Instead, successive nodes in the chain should be clustered in the same region, and the jumps across regions should be minimized: i.e., the latency for traversing the chain should be minimal (intuitively, this corresponds to solving the traveling salesman problem).

We take a simple approach to replica placement for \chainr and \name.
We start with Washington, and then traverse the continent from East to West and North to South---i.e., counter-clockwise, as follows:
(1) Washington, (2) Montreal, (3) Toronto, (4) Dallas, (5) Seattle, (6) San Jose, (7) Houston, (8) Mexico, and (9) Sao Paolo.
Each region hosts a random number of replicas between $8$ and $12$.
While this is not the optimal placement method, it is simple and yields surprisingly good results (\Cref{sec:performance-sl}).
More complex alternatives exist to our heuristic-based solution,
typically based on integer programming~\cite{wu2013spanstore}, or iterative optimization algorithms~\cite{agarwal2010volley}.

To conclude, this multi-datacenter deployment mimics a real-world global deployment, where each datacenter is effectively a city (with intra-city latencies being negligible).
This is a standard experimental setup~\cite{gilad2017algorand,rocket18avalanche,gueta2018sbft}.
Note that placing clients in Washington does not give any advantage to \chainr and \name , since each client request has to traverse the whole chain (or ring).

\paragraph{Operating system and software.}
All machines in our study run Ubuntu $14.04.1$x64.
For Apache ZooKeeper, we use $v3.4.5$~\cite{zkpkg}.
We install etcd $v2.3.7$ directly from its repository, and BFT-Smart $v1.2$~\cite{bftspkg}.

\subsection{Workload Characteristics}
\label{sec:workload}

The goal of our workload is to stress the central part of the systems under study---their underlying agreement protocol, which is typically their bottleneck.
In practice, this protocol is also in charge of replicating the request data (i.e., blocks of transactions) to all replicas.
Clients produce a workload consisting of write-only requests, i.e., we avoid read-only optimization such as leases~\cite{cha07pml}.
Each request has $250$ bytes (inspired from a Bitcoin workload~\cite{Croman16}).


\paragraph{Local Handling of Requests.}
Requests are opaque values which replicas do not interpret, consisting of a constant string.
The execution step consists of simply storing the data of each request in memory.
Recall that our main goal is to find how system size influences throughput decay.
For this reason, it is desirable to exclude features that are independent of system size, which typically would incur a fixed overhead (or amortization, in the case of batching).
Such features (e.g., persistence layer, execution)
are application-dependent, are often embarrassingly parallel, and moreover optimizations at this level~\cite{kapri12eve} are orthogonal to our study.

For ZooKeeper and etcd, we mount a \texttt{tmpfs} filesystem and configure these systems to write requests to this device.
In BFT-Smart we handle requests via a callback, which
appends every request to an in-memory linked list.
\chainr and \name simply log each request to an in-memory slice (i.e., array).

\paragraph{Batching.}
We do not optimize the batching configuration.
This is because different system sizes require different batch sizes to optimize throughput~\cite{miller2016honey}.
Moreover, batching is often an implementation detail hidden from users, e.g., etcd hardcodes the batch size to $1$MB~\cite{etcdbatching}.
Similarly, batching in ZooKeeper is not configurable; this system processes requests individually, and it is unclear whether batching is handled entirely at the underlying network layer (Netty).
In BFT-Smart we use batches of $400$, the default.
In \chainr and \name we are more conservative, allowing up to $10$ requests per batch, since these systems are already throughput-optimized in their dissemination scheme.

It is well-known that batching affects the absolute throughput of a system.
We are primarily interested, however, by the throughput \emph{decay} function of SMR systems, and not to maximize absolute throughput numbers.
Prior work has shown that batching does not affect the throughput decay, e.g., in BFT SMR systems~\cite{cowling2006hq}.
For this reason, we expect that the throughput decay in each system evolves independently of batch size.

\subsection{Workload Suite}
\label{sec:workload-suite}


Our workload suite has two parts.
We use (1) a workload \emph{generator} that creates requests and handles the communication of each client with the service.
We also use (2) a set of scripts to \emph{coordinate} the workload across all clients and control the service-side (e.g., restart service between subsequent experiments).
The workload generator differs for every SMR system, since each system has a different API.
The coordinating side is a common part which we reuse.

The main components of the workload generator are a client-side library, which abstracts over the target system, and a thread pool, e.g., using the \texttt{multiprocessing.Pool} package in Python, or \texttt{java.util.concurrent.Executors} in Java.
The thread pool instantiates parallel workload generators from a given client node.
For each of ZooKeeper, etcd, \chainr , and \name , we implement the workload generator in a Python script.
BFT-Smart is bundled with a Java client-side library; accordingly, for this system, we write the workload generator in Java.

As mentioned earlier, all client nodes are placed in the same region: Washington.
We use $10$ VMs, each hosting a client, and each client runs the workload generator instantiated with a predefined number of threads.
Depending on the target system and its size, we use between $10$ and $180$ threads per client to saturate the system and reach peak throughput.
Beside the number of threads, the workload generator accepts a few other parameters, notably, the IP and port of a system replica (this is the leader in ZooKeeper, etcd, or BFT-Smart; the head of the chain for \chainr ; or a random replica for \name); the experiment duration ($30$ seconds by default); or the size of a request ($250$ bytes in our case).

It is important that clients synchronize their actions.
For instance, client threads should coordinate to start simultaneously.
Also, we restart and clean-up the service of each system after each experiment, and we also gather statistics and logs.
The scripts for achieving this are common among all systems.
We use GNU \texttt{parallel}~\cite{tange2011gnu} and \texttt{ssh} to coordinate the actions of all the clients.
To control the service-side, we use the Upstart infrastructure~\cite{upstart}.


\section{Performance Decay Study}
\label{sec:evaluation}

We now present our observations on the performance decay of five SMR protocols.
We break this section in two parts: performance (\Cref{sec:performance-sl}) and stability results (\Cref{sec:stability-sl}).

As mentioned before, we use $10$ clients, each running multiple workload threads.
Upon connecting to a system replica, clients allow for a $20$ seconds respite to ensure all connections establish correctly.
Then all client threads begin the same workload.
Each execution runs for $30$ seconds (excluding a warm-up and cool-down time of 15 seconds each) and each point in the performance results is the average of $3$ executions.
For stability experiments we use executions of $60$ seconds, with a maximum of up to $120$ seconds.

\begin{figure*}[!ht]
\centerline{\includegraphics[width=.85\textwidth]{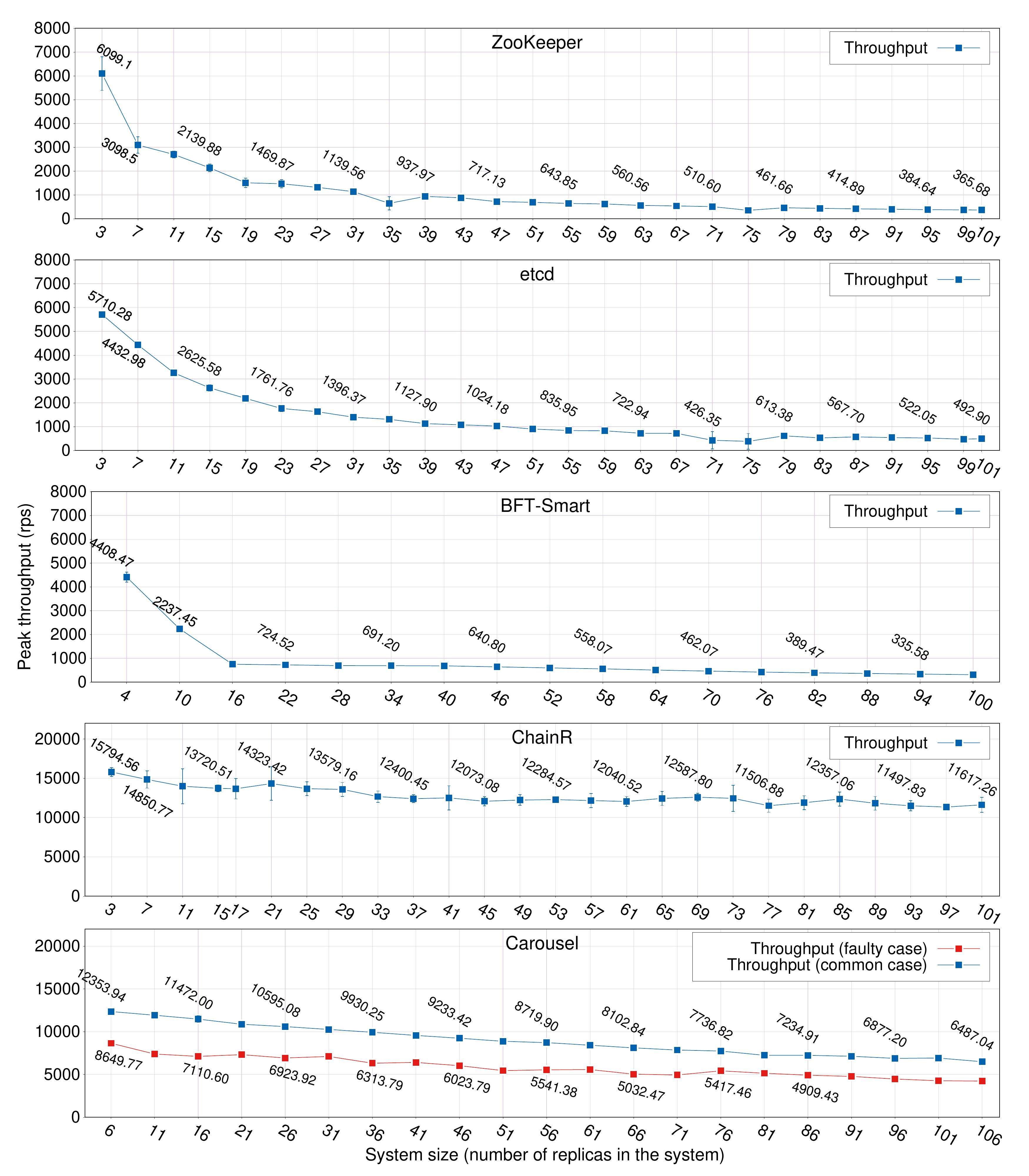}}
\caption{Throughput decay for five SMR systems on a public wide-area cloud platform. For enhanced visibility, we use separate graphs for each system.
We also indicate actual throughput values alongside data points.
Notice the different axes.}
\label{fig:scalability-1}
\end{figure*}

\subsection{Performance of SMR at 100+ Replicas}
\label{sec:performance-sl}

We first discuss throughput (\Cref{sec:performance-throughput}), and then latency (\Cref{ref:performance-sl-latency}).

\subsubsection{Throughput}
\label{sec:performance-throughput}

We report on the peak throughput value, i.e., throughput when the system begins to saturate, before latencies surge, in requests per second (\emph{rps}).
We compute this as the sum of the throughput across all $10$ clients.
We also plot the standard deviation, though often this is negligible and not visible in plots.
\Cref{fig:scalability-1} presents our results.
For readability, we also indicate throughput values on most data points.

\paragraph{ZooKeeper, etcd, and BFT-Smart.}
For ZooKeeper and etcd, which are CFT, we start from a minimum of $3$ replicas and then grow each system in increments of $4$ until we reach $101$ replicas.
For BFT-Smart, we start from $4$ replicas---the minimum configuration which offers fault-tolerance in a BFT system---and we use increments of $6$ replicas up to $N=100$.
Since we use different increments and start from different system sizes, the x-axes in~\Cref{fig:scalability-1} differ slightly across these systems.
In terms of fault thresholds, $F=\lfloor \frac{N - 1}{2}\rfloor = 50$ for CFT systems, and $F=\lfloor \frac{N - 1}{3}\rfloor=33$ for BFT-Smart.



Intuitively, there are good reasons to believe that running agreement with more than a few replicas should be avoided.
Every additional replica participates in the consensus algorithm, inflating the complexity.
This complexity, e.g., is quadratic in the number of replicas for BFT systems~\cite{cowling2006hq,Kermarrec2000}, raising the belief that SMR deployments should always stay small in their ``comfort zone.''

Prior findings on throughput decay are scarce and at small scale; briefly, these suggest that leader-centric protocols decay very sharply, e.g., up to $N\approx20$~\cite{bessani2014state,cowling2006hq,hu10zookeeper}, and consequently should be avoided.
For instance, throughput in a LAN deployment of ZooKeeper shows that each increase by $2$ in system size incurs a decay by at least $5K$ \emph{rps} (and up to $15K$)~\cite[\S5.1]{hu10zookeeper}.
The same findings are echoed in experiments with PBFT on a LAN~\cite[\S4.2]{abd05quorumupdate} or the wide-area network~\cite[\S5.2]{miller2016honey}.
Under a sustained sharp decay, throughput would drop to $0$ before reaching a few tens of replicas.

Our findings complement earlier observations at smaller scale.
Consistent with previous findings, we observe from~\Cref{fig:scalability-1} that the sharpest decline in throughput is at small system sizes.
What happens, however, is that the decay trend tapers off.
Notably from $20$ replicas onward, throughput decays more and more gracefully.


Upon closer inspection, throughput decays at a non-linear rate, roughly inversely proportional to system size $N$.
Analytically, the slopes for leader-centric systems approximate a $O(1/N)$ function.
These findings confirm an intuitive understanding of these protocols.
As discussed, ZooKeeper, etcd, and BFT-Smart are leader-centric protocols.
The leader has a certain (fixed) processing capacity, say $C$, equally divided among the other system nodes, yielding a $\sfrac{C}{N}$ decay rate.
These systems are bound by the capacity $C$ at the leader---typically CPU or bandwidth.
In our case, WAN links are not saturated; these systems are bound by leader CPU, since we use relatively small VMs~\cite{hu10zookeeper,jalil14practical}.

Throughput decay dampens as we grow each system because every additional replica incurs an amount of work depending on the current system size: adding a replica when $N{=}3$ is more costly than adding a replica when $N{=}100$, given that there are some fixed processing overheads at the primary which get amortized with system size.
We also note that in a larger system there are more tasks (such as broadcast) executing in parallel.
Finally and most importantly perhaps, throughput saturates at higher latencies when the systems are larger (see~\Cref{ref:performance-sl-latency}), since the processing pipeline depends on more replicas, i.e., as each system grows, there is a tendency to trade latency for throughput.





We observe that absolute throughput numbers at $100$ replicas are in the same ballpark for these three SMR systems, ranging from $311$ to $490$ \emph{rps}.
As a side note, this is $45-70x$ the current peak theoretical throughput of Bitcoin, suggesting that we can use SMR effectively in mid-sized blockchains, e.g., $100$ replicas.
If we extrapolate from our observation on the decay rate, it follows that these systems can match Bitcoin peak throughput at about $4500-6700$ replicas.
This is an interesting observation, as the Bitcoin network has about the same size (circa late 2016~\cite{Croman16}).
We interpret this as a simple coincidence, however.

Interestingly, BFT-Smart almost matches the performance of its CFT counterparts.
This suggests that BFT SMR protocols scale relatively as well as CFT protocols, regardless of typical quadratic communication complexity of BFT.
All three systems were surprisingly stable when running at scale, showing consistent results and consequently the standard deviations bars are often imperceptible in our plots.






\paragraph{\chainr.} We deploy \chainr using the node placement heuristic presented earlier (\Cref{sec:testbeds}).
The throughput evolution results for this system are in the fourth row of~\Cref{fig:scalability-1}; please note the y-axis range.
As expected, \chainr preserves its throughput very well with increasing system size~\cite{aublin15next700}.
So long as the system stays uniform---i.e., without reducing replica computing performance or bandwidth---the throughput degrades very slowly and at a linear rate.
At $N=101$, the system sustains $73\%$ of the throughput it can deliver when $N=3$.
This is not entirely surprising, as adding replicas to \chainr does not increase the load on any single node (in the first order of approximation), including the leader (i.e., the head replica).

To conclude, chain replication maintains throughput exceptionally well.
This system, however, sacrifices availability in the face of faults or  asynchrony.
Additionally, the chain is as efficient as its weakest link.
Indeed, we repeatedly encountered in our experiments cases with zero throughput.
Most often, this happened due to a replica crashing, but also in a few cases due to misconfiguration of a successor, therefore leaving the tail node  unreachable.

\paragraph{\name.}
This ring-based system can maintain availability despite faulty (or straggler) nodes.
Accordingly, we have two measurements: (1) a common case showing the throughput during well-behaved executions, and (2) a sub-optimal (faulty) case when $F$ faults manifest.
Since \name tolerates up to one-fifth faulty replicas~\cite{mar06fab}, we set $F=21$.
The faulty replicas occupy successive positions on the ring topology, and collaboratively they aim to create a bottleneck in the system.
They do so by activating fallback paths on the ring structure.
Concretely, they request from the same correct replica---the \emph{target}---to accept traffic from each of them and pass that traffic forward on the ring.
The target is the successor of the last faulty node.
To make matters worse, the target is also the leader replica (we call this the sequencer, described in~\Cref{sec:agreement}).
We note that faulty replicas might as well stop propagating updates; but this  has a lesser impact on throughput, as the target node would need not process the messages from all faulty nodes.
This scenario is among the worst that can happen in terms of throughput degradation to \name, barring a full-fledged DDoS attack or a crashed leader (in the latter case progress would halt in any leader-based SMR algorithm).

The results are in the fifth row of~\Cref{fig:scalability-1}.
The throughput decays similarly in the good and faulty cases; in absolute numbers there is a difference of $\mathtt{\sim}3K$ \emph{rps} at every system size.
Another way to look at it is that faulty nodes cause on average $30\%$ loss in throughput.
\name degrades less gracefully than \chainr, as it includes additional BFT mechanism (\Cref{sec:agreement}).
Nevertheless, the decay rate in \name follows a linear rate, and comes within $60\%$ of \chainr throughput (which we regard as ideal, assuming fault-free executions).
In contrast to leader-centric solutions, the chain- and ring-replication systems avoid the bottleneck at the leader and, as expected, exhibit less throughput decay, having more efficient dissemination overlays.

\subsubsection{Latency}
\label{ref:performance-sl-latency}

\begin{figure*}[t]
\centerline{\includegraphics[width=\textwidth]{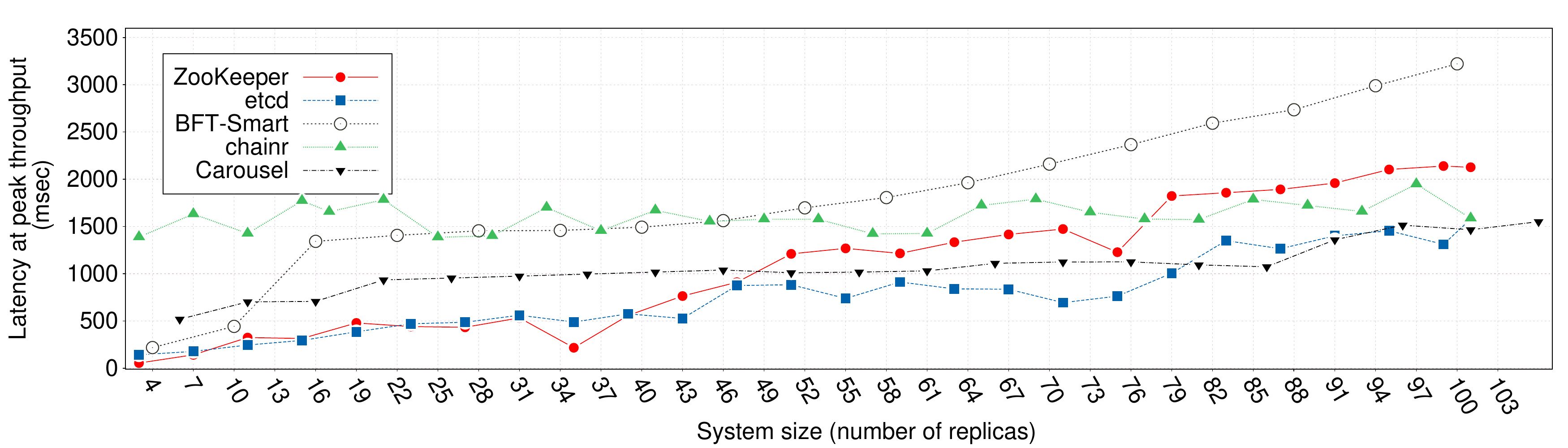}}
\caption{Average latencies (at peak throughput) for scale-out experiments with SMR systems.}
\label{fig:latencies-sl}
\vspace{-.4cm}
\end{figure*}

For latency, we report on the average latency at peak throughput, as observed by one of the clients.
Since all clients reside in Washington and connect to the same replica, they experience similar latencies.
The exception to this is in \name, where clients connect to random replicas of the system; to be fair, we present the latency of a client connecting to a replica in Washington.
The results for all five systems are in~\Cref{fig:latencies-sl}.
At small scale, CFT systems (ZooKeeper and etcd) exhibit latencies on the order of tens of milliseconds.
In contrast, BFT-Smart entails an additional phase (round-trip) in the agreement protocol for every request, which translates into higher latencies; batching, however, helps compensate for this additional phase in terms of throughput.

We remark on the high latency of \chainr.
This is to be expected, since this system trades throughput for latency, but it is also amplified by an implementation detail.
Specifically, each client runs an HTTP server, waiting for replies from the tail (a distant replica in Sao Paolo).
The server is based on the \texttt{cherrypy} framework (written in Python, and unoptimized).
At low load, the latency in \chainr is similar to \name.
But in \chainr clients create a larger volume of requests to saturate the system and also run the HTTP server, elevating the load and latency on each client. (In fact, in an earlier version of \chainr, clients were the bottleneck.)

The average latency across all SMR systems does not surpass $3.5$ seconds.
We only include the latency for the good case of \name, but even in the faulty case, latency does not exceed $3.2$s.
In terms of 99th percentiles, the worst cases are $6.5$s for BFT-Smart ($N=100$), and $6$s for the faulty-case of \name ($N=107$).

\subsection{Stability experiments}
\label{sec:stability-sl}

Our goal here is to evaluate if mature SMR systems recover efficiently from a serious fault when running at large scale.
Concretely, we crash the leader replica (triggering the leader-change protocol) and measure the impact this has on throughput.
We cover ZooKeeper, etcd, and BFT-Smart; the other systems (\chainr and \name) have no recovery implemented.

\begin{figure}[t]
\centerline{\includegraphics[width=0.8\columnwidth]{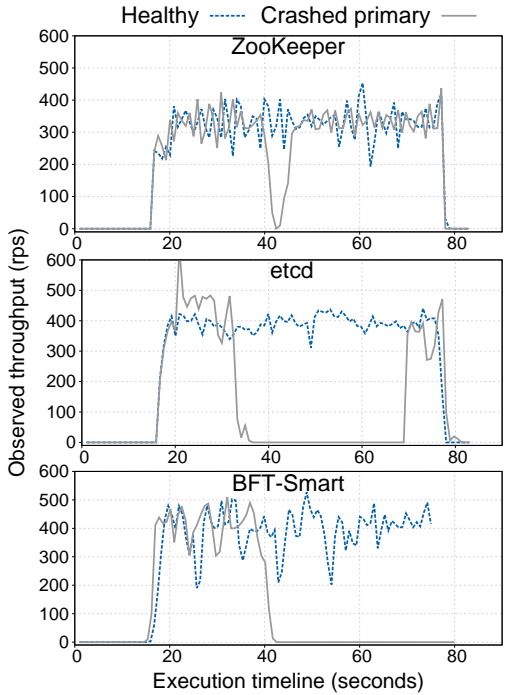}}
\caption{Stability experiments for ZooKeeper and etcd ($N=101$), as well as for BFT-Smart ($N=100$).}
\label{fig:stability}
\vspace{-.4cm}
\end{figure}


When the leader in an SMR system crashes, this kicks off an election algorithm to choose a new leader.
We study the stability of this algorithm---whether it works at scale and how much time it requires.
These tests are rather about the code maturity in these systems rather than their algorithmic advantages.
Our results are in~\Cref{fig:stability}, describing two runs: (1) a \emph{healthy} case, and (2) a case where we \emph{crash} the leader (i.e., primary).
During the first $20$s clients simply wait, and then they start their workload; we crash the leader $20$s later.
The point where we crash the leader is obvious, around $40$s, as the throughput drops instantly to zero.

ZooKeeper has consistently the fastest recovery.
The throughput reaches its peak within just a few seconds after the leader crashes, consistent with earlier findings at smaller scale~\cite{shraer12dynamic}.
This system has an optimization so that the new leader is a replica with the most up-to-date state, which partly accounts for fast recovery~\cite{ongaro2014consensus}.
In etcd recovery is slower: It can take up to $40$ seconds for throughput to return to its peak.
The election mechanism in etcd is similar to that of ZooKeeper~\cite{ongaro2014consensus}, and the heartbeat parameters of these two systems are similar as well ($1$ and $2$ seconds, respectively).
The difference in stability between ZooKeeper and etcd can also stem from an engineering aspect, as the former system has a more stable codebase (started in 2007) compared with the latter system (2013).

For BFT-Smart, we allowed the system up to $120$ seconds to recover but throughput remained $0$.
We also tried with smaller sizes, and found that $N=88$ is the largest size where BFT-Smart manages to recover, after roughly $10s$.
Finally, we remark that we were able to reproduce all these behaviors across multiple (at least $3$) runs, so these are not outlying cases.

\section{\name Protocol}
\label{sec:common-case}

An interesting design aspect of \name is how the ring overlay masks faults.
We achieve this by keeping redundant paths on this overlay, thus ensuring availability despite faults or asynchrony.
The agreement protocol of \name is of interest as well, which is the FaB consensus algorithm~\cite{mar06fab} adapted to a ring overlay.


We choose to pattern the agreement protocol of \name after FaB~\cite{mar06fab} 2-step BFT consensus algorithm due to the interesting tradeoff this offers.
FaB reduces the latency of BFT agreement from three steps to two.
This is appealing in our ring topology, because each step is a complete traversal of the ring.
Fewer traversals means higher throughput, lower latency, and a simpler protocol than 3-step ones~\cite{castro2002}.
This benefit comes to the detriment of resilience: the system needs larger quorums to tolerate faults.
\name assumes $N=5F+1$ replicas, whereas optimal BFT systems tolerate one-third faults, i.e., $N=3F+1$~\cite{castro2002}.
As in prior solutions, our agreement protocol relies on the existence of a \emph{sequencer} (i.e., a leader) assigning sequence numbers to operations.

\begin{figure}[t]
  \centerline{\includegraphics[width=.65\columnwidth]{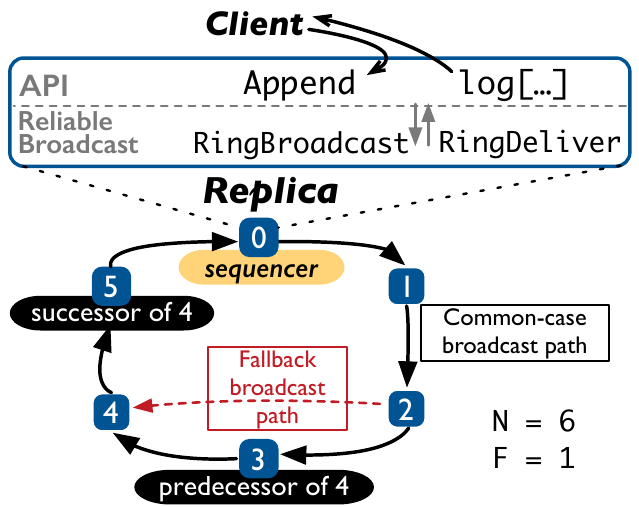}}
  \caption{Overview of \name in a system of $6$ replicas. Clients interact with the system via a thin API. Underlying this API, there is a reliable broadcast scheme executing along a ring overlay network. In this overlay, each replica has a \emph{successor} and \emph{predecessor}. There is a specific replica which carries the role of a \emph{sequencer} (replica $0$ here).}
  \label{fig:logical-layering}
  \vspace{-.4cm}
\end{figure}

\Cref{fig:logical-layering} shows an overview of \name when $N=6$.
The replicas, labeled $0...5$, are organized on a ring.
Note that each replica in this overlay has a certain \emph{successor} and \emph{predecessor} replica.
One of the replicas (node $0$) is the sequencer.
By default, broadcast messages disseminate through the system from one replica to its immediate successor.
Additionally, fallback (redundant) paths exist, to ensure availability despite asynchrony or faults.

The interface of \name is log-based: Each replica exposes a simple API which allows clients to read from a totally-ordered \emph{log} and to \emph{Append} new entries to this log.
Under this API layer, all replicas implement a reliable broadcast primitive providing high throughput and availability.
We discuss the Append operation first (\Cref{sec:agreement}), then the reliable broadcast layer (\Cref{sec:reliable-protocol}), followed by the reconfiguration sub-protocol (\Cref{sec:reconfiguration}) and correctness arguments (\Cref{sec:correctness}).

\subsection{Append Operation}
\label{sec:agreement}

To add an entry \emph{e} to the totally-ordered log, clients invoke Append(\emph{e}) at any replica $i$.
The operation proceeds in two logical phases:
\begin{compactenum}
    \item \textbf{Data}---Replica $i$ broadcasts entry \emph{e} to all correct replicas using \emph{RingBroadcast} of the underlying broadcast layer.
    \item \textbf{Agreement}---A BFT agreement protocol executes. The sequencer proposes a sequence number (i.e., a log position) for entry \emph{e}, and correct replicas confirm this proposal. After executing the agreement phase for this entry, replica \emph{i} notifies the client that the operation succeeded.
\end{compactenum}

\Cref{lst:total-order-protocol} shows the implementation of the Append operation.
First, replica $i$ broadcasts a $\langle\opname{data}, e\rangle$ message, as shown on line~\ref{line:disseminate}.
This corresponds to the first logical phase of the Append operation.
We say that this broadcast message is of \emph{type data} and has \emph{payload} \emph{e}.
As this message disseminates throughout the system, each correct replica triggers the \emph{RingDeliver} callback to deliver the data message (line~\ref{line:callback} of~\Cref{lst:total-order-protocol}).

The delivery callback always provides two arguments: (1) an identifier \emph{id}, and (2) the actual message.
The underlying broadcast layer assigns the \emph{id}, which uniquely identifies the associated message.
We discuss identifiers in further detail later, but suffice to say that an identifier is a pair denoting the replica which sent the corresponding message plus a logical timestamp for that replica (\Cref{sec:reliable-protocol}).
The actual message, in this case, is a data message with payload entry~\emph{e}.
Upon delivery of any data message, each correct replica stores this entry in a \emph{pending} set.
Note that this set is indexed by the assigned \emph{id} (line~\ref{line:pending}).

The agreement phase starts when the sequencer replica delivers the data message with~\emph{e}.
After saving \emph{e} in the pending set, the sequencer also proposes a sequence number for \emph{e} by broadcasting a
$\langle\opname{AGREEMENT}, sn, id, hash\rangle$ message (line~\ref{line:send-seq-1} and lines~\ref{line:send-seq-2}--\ref{line:send-seq-3}).
It would be wasteful (in terms of bandwidth) to include the whole entry in this agreement message; instead, the sequencer simply pairs the entry \emph{id} with a monotonically increasing sequence number \emph{sn} (called \emph{nextSeqNr} on line~\ref{line:send-seq-3}).
The \emph{hash} in the agreement message is computed on the concatenation of the assigned sequence number, the entry \emph{id}, and the entry \emph{e} itself.

\begin{lstlisting}[mathescape,
    float,
    floatplacement=Hb,
    language=Carousel,
    caption={A high-level algorithm describing the Append operation.},
    label={lst:total-order-protocol}]
func Append(e):
  RingBroadcast(<DATA, e>) // Broadcast data message with payload 'e' %*\label{line:disseminate}*)

callback RingDeliver(id, <type, payload>): %*\label{line:callback}*)
  if (type == DATA):
    pending[id].entry = payload %*\label{line:pending}*)
    proposeAgreementMsg(id, payload) // Executes ONLY at sequencer %*\label{line:send-seq-1}*)
  else if (type == AGREEMENT):
    handleAgreementMsg(payload)  // The payload is a triplet

// Executes ONLY at the sequencer
func proposeAgreementMsg(id, e): %*\label{line:send-seq-2}*)
  hash = Hash(nextSeqNr . id . e)
  // Disseminate the agreement message
  RingBroadcast(<AGREEMENT, nextSeqNr++, id, hash>) %*\label{line:send-seq-3}*)

func handleAgreementMsg(sn, id, hash):%*\label{line:pending-1}*)
  validate(sn, id, hash) || return%*\label{line:validate}*)
  pending[id].sn   = sn %*\label{line:indexing}*)
  pending[id].hash = hash%*\label{line:pending-2}*)
  pending[id].confirmations++
  // Disseminate our own confirmation for this sequence number
  RingBroadcast(<AGREEMENT, sn, id, hash>) %*\label{line:send-seq-4}*)
  if (pending[id].confirmations == 4F+1):
    addToStableLog(id) %*\label{line:stable}*)
\end{lstlisting}

Each replica delivers the agreement message of the sequencer trough the same RingDeliver callback of the broadcast layer.
After any replica $j$ delivers the agreement message, then $j$ broadcasts its own agreement message with the same triplet (\emph{sn, id, hash}), as described on line~\ref{line:send-seq-4}.
Prior to doing so, replica~$j$ validates the hash and saves the assigned sequence number and hash in the pending log (lines~\ref{line:pending-1}--\ref{line:pending-2}).

The validation step (line~\ref{line:validate}) refers to several important checks: that the \emph{hash} correctly matches the sequence number \emph{sn}, identifier \emph{id}, and entry content; that this \emph{id} has no other proposal for a prior sequence number; that each number \emph{sn} in an agreement message has a corresponding proposal for that \emph{sn} from the sequencer replica; and that different agreement messages confirming the same \emph{sn} originate from distinct replicas.

We say that a replica commits on log entry \emph{e} after it gathers sufficient ($4F+1$) confirmations for that entry.
The entry then becomes stable at that replica (line~\ref{line:stable}).
By the reliability of the RingBroadcast dissemination primitive, if the sequencer is correct and proposes a valid $sn$, then the entry eventually becomes stable at all correct replicas.
As proved in prior work~\cite{mar06fab}, this protocol has optimal resilience for two-step BFT agreement, i.e., $N=5F+1$ is the smallest system size to tolerate $F$ faults with two-step agreement.

\begin{figure}[t]
  \centerline{\includegraphics[width=.75\columnwidth]{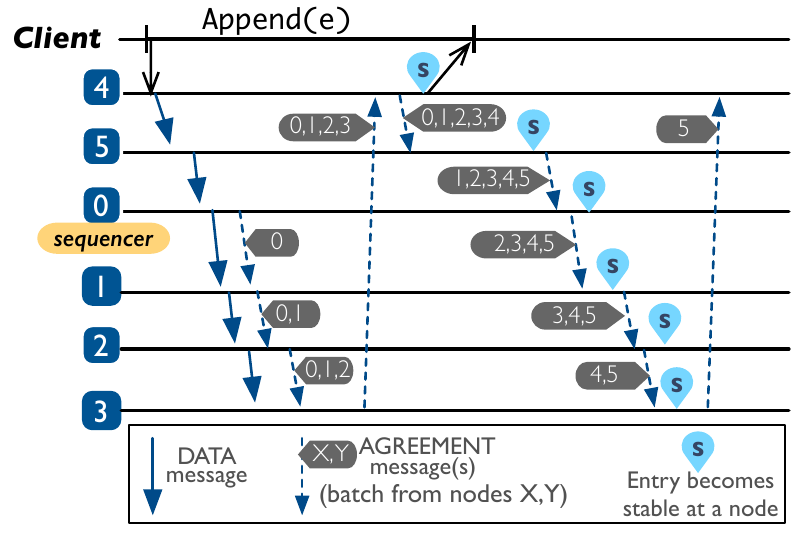}}
  \caption{The unfolding of an Append(\emph{e}) operation in \name. The client contacts replica $4$, which broadcasts a data message. Once this message reaches the sequencer (replica $0$), this replica broadcasts an agreement message. Then all replicas broadcast their agreement message. Entry \emph{e} becomes stable once a replica gathers at least $4F+1=5$ matching agreement messages for \emph{e}.}
  \label{fig:msg-lifecycle}
\end{figure}

Informally, as the initial agreement message (from the sequencer) propagates on the ring, it produces a snowball effect: Each replica delivering this message broadcasts its own agreement message, confirming the proposed sequence number.
\Cref{fig:msg-lifecycle} depicts this intuition.
Notice that the data and agreement phases partially overlap (beginning from replica $0$).
Overall, it takes $N{+}2$ message delays for replica $4$ to respond to the client request.
In terms of message complexity, each replica processes $N{+}1$ messages.
In other words, there are $N{+}1$ total invocations to RingBroadcast: one with the data message, plus $N$ agreement messages (one per replica).

\subsection{Reliable Broadcast in a Ring Topology}
\label{sec:reliable-protocol}

In a conventional ring-based broadcast, each replica $i$ expects its predecessor $i\textrm{-}1$ to forward each message which $i\textrm{-}1$ delivers~\cite{guerraoui10ring,jalili17ring}.
Messages travel from every replica to that replica's successor.
Intuitively, this scheme is throughput-optimal because it balances the burden of data dissemination across all replicas~\cite{guerraoui10ring}; the downside, however, is that asynchrony (or fault) at \emph{any} replica can affect availability by impeding progress.
To maintain high-throughput broadcast across the ring overlay despite asynchrony or faults, in \name we strengthen the ring so that each replica connects with $F+1$ total predecessors.
A replica has one \emph{default} connection---with the immediate predecessor---and up to $F$ \emph{fallback} connections---with increasingly distant predecessors.
The topology we obtain is essentially an \mbox{$F$-connected} graph.
This graph ensures connectedness (availability) despite up to $F$ faults.

In~\Cref{fig:logical-layering} for instance, replica $4$ should obtain from replica $3$ all messages circulating in the system.
If replica $3$ disrupts dissemination and drops messages, however, then replica $4$ can activate the fallback connection to replica $2$.
By default, communication on this fallback path is restricted to brief messages called \emph{state vectors}, which replica $2$ periodically sends directly to replica $4$.

More generally, replica $i$ expects a state vector from all $F$ of its fallback predecessors, i.e., from replicas $i\textrm{-}2, i\textrm{-}3,...$.
A state vector is a concise representation of all messages delivered by the corresponding fallback replica.
If replica $i$ notices that its immediate predecessor $i\textrm{-}1$ is omitting messages (and that the state on fallback replica $i\textrm{-}2$ is steadily growing), then $i$ sends a $\langle\opname{ACTIVATE}, sv_i\rangle$ message directly to replica $i\textrm{-}2$, where $sv_i$ is the state vector of replica $i$.

Replica $i\textrm{-}2$ interprets the activate message by sending to replica $i$ all messages which $i\textrm{-}2$ has delivered and are not part of state vector $sv_i$.
Thereafter, replica $i\textrm{-}2$ continues sending to $i$ any new messages it delivers.
In the meantime, replica $i\textrm{-}2$ also continues forwarding messages as usual to its immediate successor replica $i\textrm{-}1$.
In case replica $i\textrm{-}1$ restarts acting correctly and forwards messages to $i$, then replica $i$ can send a $\langle\opname{DEACTIVATE}\rangle$ message to $i\textrm{-}2$.

Alternatively, replica $i$ can request individual pieces of the state from $i\textrm{-}2$, e.g., in case replica $i\textrm{-}1$ is selectively withholding messages from $i$.
Since every replica has $F+1$ total connections, \name can tolerate up to $F$ faults, regardless whether these faults are successive on the ring or dispersed across the system.
This mechanism based on fallback connections is strictly to improve availability (i.e., delivery of broadcast messages) relying on timeouts, but does not affect the safety of the protocol (\Cref{sec:correctness}).

The state vector $sv_i$ at some replica $i$ is a vector of timestamps with one element per replica.
The element on position $j$ denotes the latest message broadcast by replica $j$ which replica $i$ delivered.
Concretely, each such element is a timestamp, i.e., a logical counter attached to any message which a replica sends upon invoking RingBroadcast.
For instance, whenever replica $j$ calls RingBroadcast(\emph{m}), the broadcast layer tags message $m$ with a unique \emph{id}, in the form of a pair $\{j, ts\}$, where $j$ denotes the sender replica and $ts$ is a monotonically increasing timestamp specific to replica $j$.~\footnote{Just like sequence numbers, timestamps are \emph{dense}~\cite{castro2002}: This prevents replicas from exhausting the space of these numbers and makes communication steps more predictable, which simplifies dealing with faulty behavior~\cite{aiyer05bar}.}
As we explained earlier, the \emph{id} also plays an important role in the Append operation (\Cref{sec:agreement}).

When replica $i$ delivers message $m$ with id $\{j, ts\}$ from replica $j$ via the RingDeliver callback, replica $i$ updates its state vector $sv_i$ to reflect timestamp $ts$ for position \emph{j}.
Status vectors are inspired from vector clocks~\cite{fid88times,matt88virtual}.
The notable difference to vector clocks is that a replica does not increment its own timestamp when delivering or forwarding a message: This timestamp increments only when a replica sends a new message---typically a data or a agreement message---by invoking RingBroadcast.
With state vectors, our goal is not to track causality or impose an order~\cite{guerraoui10ring}, but to ensure no messages are lost (i.e., reliability).


One corner-case that can appear in our protocol is when a malicious sender replica attempts to stir confusion using incorrect timestamps.
In particular, such a replica $i$ can attach the same timestamp to two different messages: $m1$ with $\{i, ts1\}$ and $m2$ with id $\{i, ts1\}$.
Another bad pattern is skipping a step in the timestamp by broadcasting first $\{i, ts1\}$ and then $\{i, ts3\}$.
In practice, communication between any two replicas relies on FIFO links (e.g., TCP), so correct replicas can simply disallow---as a rule---gaps or duplicates in messages they deliver.

FIFO links, however, do not entirely fix the earlier problem.
It is possible that some replica delivers $m1$ with timestamp $ts1$, while a different replica delivers $m2$ for this same timestamp.
But note that this will not cause safety issues.
The replicas can only agree on either of $m1$ or $m2$: When the sequencer proposes a sequence number for id $\{i, ts1\}$, it also includes a hash of the corresponding message, either $m1$ or $m2$.
Even if the sequencer is incorrect and proposes sequence numbers for \emph{both} $m1$ and $m2$ (a poisonous write~\cite{mar06fab}) only one of these two messages can gather a quorum and become stable.
To conclude, timestamps restrict acceptable behavior and---when coupled with the sender replica's identity---provide a unique identifier for all messages circulating in the system, which helps ensure reliability.

\subsection{Reconfiguration}
\label{sec:reconfiguration}

The reconfiguration sub-protocol in \name ensures liveness when the sequencer replica misbehaves.
This protocol is, informally, an adaptation of the FaB~\cite{mar06fab} recovery mechanism (which is designed for an individual instance of consensus) to state machine replication.

\subsubsection{Preliminaries}

Reconfiguration concerns the agreement algorithm in \name.
Neither the data phase (of the Append operation) nor the broadcast layer need to change.
We adjust the common-case protocol of~\Cref{sec:agreement} to accommodate reconfiguration as follows:
\begin{compactenum}
\item Replicas no longer agree on a sequence number $sn$ for every Append operation.
Instead, each instance of the agreement protocol for a given sequence number is tagged with a \emph{configuration number}, so that replicas now agree on a tuple $\{\propn, sn\}$.
In other words, agreement messages now have the form $\langle\opname{AGREEMENT}, \{\propn, sn\}, id, hash\rangle$.
The concept of \emph{proposal numbers} in FaB~\cite{mar06fab} or that of \emph{view numbers} in PBFT~\cite{castro2002} is analogous to configuration numbers in \name ; briefly, these serve the purpose of tracking the number of times the sequencer changes.

\item We modify the $hash$ in agreement messages to also include the configuration number.

\item Our common-case protocol assumes that, for every sequence number $sn$, a replica only ever accepts (i.e., gives its vote for) a single agreement message, namely, the first valid agreement message they deliver for that $sn$ from the sequencer.
To account for configuration numbers, a replica is now allowed to change its mind, and accept another agreement message if the sequencer changed (i.e., in a different configuration number).

\end{compactenum}

Configuration numbers in \name start from $0$.
Every time the configuration number increases, the sequencer role changes deterministically, so that the new sequencer is the successor of the previous sequencer, in a round robin manner.

We note that the ring-based dissemination algorithm (\Cref{sec:reliable-protocol}) that replicas employ during common-case requires no modifications.
While executing reconfiguration (described next), however, replicas do not use the ring-based broadcast primitive.
If reconfiguration is executing, this means that the current sequencer is faulty and hence the system is experiencing no progress.
For this reason, we can temporarily renounce on the high-throughput broadcast enabled by the ring topology, and adopt instead a conventional all-to-all broadcast scheme towards optimizing for latency.

\subsubsection{Protocol}
\label{sec:reconfiguration-protocol}

A correct replica $i$ enters the reconfiguration sub-protocol if any of the following conditions hold: (1) a timer expires at replica $i$ because the sequencer replica failed to create new agreement messages or a previous reconfiguration failed to complete in a timely manner, or (2) replica $i$ observes $F+1$ other replicas proposing reconfiguration.

To propose a reconfiguration and change the sequencer, replica $i$ broadcasts a \reconfigg{i}{\propn} message.
Once it does so, replica $i$ also starts ignoring any messages concerning
configuration number $\propn$ or lower, and until reconfiguration completes it ignores any messages except those of type data, reconfiguration, or new-configuration (as we define them below).
Replica $i$ also starts a timer to prevent a stalling reconfiguration.

The set \pendinglog{} in a reconfiguration message contains agreement messages which replica $i$ signed for all the sequence numbers which are still pending at this replica.
In other words, these are proposals for entries which are not part of the stable log at replica $i$ because this replica gathered insufficient confirmations (i.e., less than $4F+1$) to mark the corresponding entry as stable.

The successor of the faulty (old) sequencer, namely the replica at position $\propn'$ on the ring, where $\propn' = (\propn+1) \% N$, is set to become sequencer when reconfiguration completes.
This new sequencer waits until it gathers reconfiguration messages from $4F+1$ replicas, including itself, and then broadcasts a \newconfigg\ message.
Here, \allreconfig\ represents the set of $4F+1$ reconfiguration messages which the new sequencer gathered.

Reconfiguration completes at replica $i$ when $i$ delivers the new-configuration message.
When this happens, replica $i$ starts accepting agreement messages created by the new sequencer (i.e., the replica at position $\propn'$ on the ring) and expects these messages to be tagged with configuration number $\propn+1$.
After reconfiguration completes, the new sequencer starts redoing the common-case agreement protocol for every individual sequence number found in an agreement message in the set \allreconfig .

For every sequence number $sn$ appearing in \allreconfig , the new sequencer creates a new agreement message with the configuration number set to $\propn+1$.
When creating these new agreement messages, for every sequence number $sn$ there are two cases to consider:

\emph{(1)} If there exists an $id$ that appears in an agreement message of \allreconfig,
such that there is no $id' \neq id$ with $2F+1$ agreement messages (with the same $sn$) in \allreconfig,
then the new sequencer chooses this $id$ to be associated to $sn$ in the new agreement message.
The new sequencer also recomputes the hash as done in the common-case protocol, including the new configuration number $\propn+1$.
In FaB terminology~\cite{mar06fab}, we say that the set \allreconfig\ of reconfiguration messages \emph{vouches} for $id$ to be associated to $sn$.

\emph{(2)} Alternatively, it can happen that more than one pair $(sn,id)$ is vouched for by \allreconfig{} (or no such pair at all).
This can occur, for instance, if the previous sequencer was malicious and proposed for the same sequence number $sn$ multiple different $id$s.
In this case, the sequencer can choose to associate $sn$ with any $id$ that was previously proposed for $sn$, and recomputes the hash as done in the common-case.
For every such new agreement message that the new sequencer creates, the common-case protocol executes as described in~\Cref{sec:agreement}, accounting for the configuration number $\propn+1$.

\subsection{Correctness Arguments}
\label{sec:correctness}

\noindent\textbf{Safety.}
\name provides three safety guarantees.
  (1) Only values (i.e., entry $id$s) that are proposed by replicas can become stable;
  (2) Per sequence number, only one value can become stable;
  (3) The same value cannot become stable at two different sequence numbers.

Recall that the agreement algorithm in \name is patterned after the FaB protocol.
Specifically, we rely on the correctness of this protocol, so properties (1) and (2) follow from the safety properties of FaB (CS1 and CS2, respectively)~\cite[\S5.4]{mar06fab}.

Property (3) follows from the fact that correct replicas index values by their identifier \emph{id} (see lines~\ref{line:pending} and~\ref{line:indexing} in~\Cref{lst:total-order-protocol}); hence each value will be associated to a unique sequence number, and no correct replica will accept or confirm any value that was previously proposed for a different sequence number.

\noindent\textbf{Liveness.}
Our system ensures the following liveness properties.
  (1) at any point in time, if there exists a value proposed by a correct
    replica which is not part of the stable log yet, then eventually one value
    (possibly another one) will become stable,
  (2) if there are only a finite number of proposed values, then all of them
    eventually become stable.

  (1) Assume there is a value $v$ (entry id) which is not part of the stable log yet.
  Let $\seqn$ be the first sequence number for which no value was accepted.
  If the current sequencer is correct, then it will propose a unique
  value for sequence number $\seqn$ (either value $v$, or another one), and the
  replicas will reach agreement on that value.

  If the replicas are not able to eventually reach agreement for sequence number $\seqn$,
  this means that the current sequencer is faulty, and they will trigger a
  reconfiguration. The new sequencer might be faulty as well, but after
  enough reconfigurations (at most $F$), a correct sequencer is chosen.
  This new sequencer will then propose a unique value for sequence number
  $\seqn$ that will become accepted by all (correct) replicas, and accepted
  to the stable log.

  (2) We apply the first liveness property as many times as there are proposed
  values. Together with safety, this ensures that every operation that needs to
  be ordered eventually becomes stable.

\section{Discussion \& Related Work}
\label{sec:discussion-related}


Several strongly-consistent systems, including those based on chain- or ring-replication, have been designed assuming a synchronous system model with uniform replica and network characteristics~\cite{aublin15next700,knezevic12high,van04chain}.
By using an efficient communication pattern (i.e, load-balancing the agreement algorithm) these systems exhibit the best performance decay.
These protocols seem to excel in synchronous models~\cite{guerraoui10ring}---and forfeit availability otherwise.
In a WAN (or multi-datacenter), uniformity or synchrony is unlikely.
We introduce the \name protocol as a first step in trying to bridge the gap between good performance decay and good availability.


We remark on two concrete research directions to help further the goal of reconciling performance decay and availability in agreement protocols.
First, it is appealing to combine broadcast with ring-based protocols in a single system, in the vein of Abstract~\cite{aublin15next700}.
Such a system could provide higher BFT resilience, namely tolerate $F=\lfloor (N - 1)/ 3\rfloor$ faults, unlike \name where resilience is $F=\lfloor (N - 1)/ 5\rfloor$.
Combining protocols typically yields intricate systems, however, and it is important to address the resulting complexity.
Second, for predictable behavior, faulty nodes should be detected and evicted from the system in a timely manner.
This is challenging in a Byzantine environment.
Past approaches rely on proofs-of-misbehavior (which are useful for limited kinds of faults~\cite{kot07zyz}) or on incentives (which tend to be complex~\cite{aiyer05bar}), so new solutions or practical assumptions are needed.




Chain- or ring-replication is a well-studied scheme in SMR.
Our baseline, \chainr, is a straightforward implementation of chain replication in the fail-stop model, shedding light on what is an ideal upper bound of SMR throughput in fault-free executions.
In contrast to \name , previous approaches consider a model that does not tolerate Byzantine faults~\cite{amir05spread,and09fawn,jalili17ring}, or they degrade to a broadcast algorithm to cope with such faults~\cite{aublin15next700,knezevic12high}.

The technique of overlapping groups of chain replication on a ring topology, as employed in FAWN~\cite{and09fawn}, is similar to basic ring-based replication.
The insight is similar: instead of absorbing all client operations through a single node (a bottleneck), accept operations at multiple nodes.
The most important distinction between FAWN and \name lies in the use of  sharding.
FAWN shards the application state, each shard mapping to a chain replication group.
As we mentioned earlier, sharding is a common workaround to scale SMR~\cite{bezerra2016strong,glenden11scatter,co13spanner}.
In \name and other systems we study, the goal is \emph{full} replication.
Even when sharding is employed, our findings are valuable because they apply to the intra-shard protocol (which is typically an SMR instance).


S-Paxos~\cite{bi12spaxos} decouples request \emph{dissemination} from request \emph{ordering}.
This relaxes the load at leader and can increase throughput.
We apply this same principle in \name , by separating dissemination (ring-based broadcast for high throughput) from agreement (FaB).
In contrast to \name, S-Paxos does not tolerate Byzantine faults.

Building upon a conjecture of Lamport~\cite{lamp03lowerbounds}, FaB~\cite{mar06fab} laid the fundamental groundwork for 2-step BFT consensus.
The agreement algorithm in \name is a simplified FaB protocol (e.g., we use only one type of protocol message, $\opname{SEQUENCE}$, for reaching agreement), but the most important distinction to FaB is that \name employs a ring topology in the common-case, eschewing the throughput bottleneck at the leader.
We believe the FaB agreement algorithm combined with dissemination schemes from the chain- or ring-based families~\cite{guerraoui10ring,jalili17ring,van04chain} deserve more attention.

It was recently uncovered that a version of the FaB protocol is flawed.
Namely, that this protocol version suffers from liveness issues, which can happen when a malicious leader engages in a poisonous write~\cite{abrah17revisiting}.
This problem, however, only applies specifically to the parametrized version of FaB~\cite{mar06fab}, called PFaB in~\cite{abrah17revisiting}.
PFaB is not the same protocol as the one we use in \name, hence \name is not subject to the liveness problems of PFaB.

It is difficult---if not impossible---to be comprehensive in studying agreement protocols, given the abundance of work on this topic.
In choosing the five systems in this paper, our goal was to include solutions that are as diverse as possible in their design while also being broadly applicable.
Many interesting agreement protocols exist that build on various assumptions to speed performance and allay their throughput decay, for instance by relying on correct client behavior as in Zyzzyva~\cite{kot07zyz}, exploit application semantics as in EPaxos~\cite{mora13epaxos}, considering certain restricted data types as in the Q/U protocol~\cite{abd05quorumupdate}, or client speculation~\cite{weste09latency}.


Our stability experiments show that BFT SMR protocols, even as mature as BFT-Smart, are not as battle-tested as CFT protocols.
Indeed, open-source implementations of BFT SMR are scarce, and the issue of \emph{non-fault-tolerant} BFT protocols is known~\cite{clement09making}.
In fact, it is a pleasant surprise for us that BFT-Smart is able to go through reconfiguration at $88$ replicas (despite folding at bigger system sizes), and resume request execution after the leader fails.

An important application of Byzantine fault-tolerant SMR is for permissioned distributed ledger applications~\cite{hyperledger,sou18byzantine}.
In such an application, SMR can serve as an essential sub-system ensuring a total-order across the entries in the ledger.
There is a growing body of work dealing with the problem of scaling consensus for distributed ledgers, which we cover briefly.

SBFT is a recent BFT system showing impressive performance, e.g., at least $350$ \emph{rps} when $F{=}64$ (i.e., $N{=}193$)~\cite{gueta2018sbft}. To obtain a scalable solution, this system explores modern cryptographic tools (plus other optimizations), a design dimension we did not explore in this work.
Another practical method to scale agreement to a large set of nodes is to elect a small committee (or even one node), and run the agreement protocol in this committee.
This method appears in Algorand~\cite{gilad2017algorand}, HoneyBadger~\cite{miller2016honey}, ByzCoin~\cite{kogi16byzcoin}, or Bitcoin-NG~\cite{eyal16bitcoinng}, among others.
Most research in this area seeks to ensure probabilistic guarantees, whereas we consider deterministic solutions.
Nevertheless, our findings in this paper apply to most of these protocols, because typically the committees in these systems run a deterministic protocol, e.g., PBFT~\cite{castro2002,kogi16byzcoin}.



A specific step that is common to all SMR protocols is the processing of protocol messages and client requests at each replica.
Several approaches can optimize this step.
These are all orthogonal to our study, and they generally apply to any SMR system.
Examples include optimistic execution to leverage multi-cores~\cite{kapri12eve} or hardware-assisted solutions~\cite{poke15dare}, e.g., to speed costly crypto computations, offload protocol processing~\cite{istva16consbox}, or using a trusted module to increase resilience and simplify protocol design~\cite{chun07attested}.

\section{Conclusions}
\label{sec:conclusions}

It is commonly believed that throughput in agreement protocols should degrade sharply with system size.
The data supporting this belief is scarce, however, and simple extrapolation cannot tell the whole story.
Consequently, in this paper we have studied empirically the performance decay of agreement in five SMR systems.



A positive takeaway from our study was that mature SMR implementations (ZooKeeper, etcd, and BFT-Smart) can sustain, out-of-the-box, $300$--$500$ \emph{rps} (requests per second) at $100$ replicas.
Their throughput decays rapidly at small scale (consistent with previous findings), but this decay dampens at larger size.
These systems exhibit a non-linear (inversely proportional) decay rate.
Throughput decays very slowly (linearly) in chain-replication, and this system can sustain $11K$ \emph{rps} at $100$ replicas.
\name comes within $60\%$ performance of the ideal (chain-replication).
This suggests that
SMR can be effectively employed in mid-sized applications, e.g., up to hundreds of replicas, and there should be more focus on chain- or ring-based topologies for mitigating agreement protocols performance decay.




\raggedright

\bibliographystyle{acm}
\bibliography{refs}

\begin{thebibliography}{10}

\bibitem{abd05quorumupdate}
{\sc Abd-El-Malek, M., Ganger, G.~R., Goodson, G.~R., Reiter, M.~K., and Wylie,
  J.~J.}
\newblock {Fault-scalable Byzantine Fault-tolerant Services}.
\newblock {\em SIGOPS Oper. Syst. Rev. 39}, 5 (2005), 59--74.

\bibitem{abrah17revisiting}
{\sc Abraham, I., Gueta, G., Malkhi, D., Alvisi, L., Kotla, R., and Martin,
  J.-P.}
\newblock Revisiting fast practical byzantine fault tolerance.
\newblock {\em arxiv:2017\/} (2017).

\bibitem{ab13sharding}
{\sc Abu-Libdeh, H., van Renesse, R., and Vigfusson, Y.}
\newblock {Leveraging Sharding in the Design of Scalable Replication
  Protocols}.
\newblock In {\em SOCC\/} (2013).

\bibitem{adya2010centrifuge}
{\sc Adya, A., Dunagan, J., and Wolman, A.}
\newblock Centrifuge: Integrated lease management and partitioning for cloud
  services.
\newblock In {\em NSDI\/} (2010).

\bibitem{agarwal2010volley}
{\sc Agarwal, S., Dunagan, J., Jain, N., Saroiu, S., Wolman, A., and Bhogan,
  H.}
\newblock Volley: Automated data placement for geo-distributed cloud services.
\newblock In {\em NSDI\/} (2010).

\bibitem{agu11das}
{\sc Aguilera, M.~K., Keidar, I., Malkhi, D., and Shraer, A.}
\newblock Dynamic atomic storage without consensus.
\newblock {\em J. ACM 58}, 2 (2011), 7:1--7:32.

\bibitem{aiyer05bar}
{\sc Aiyer, A.~S., Alvisi, L., Clement, A., Dahlin, M., Martin, J.-P., and
  Porth, C.}
\newblock {BAR} fault tolerance for cooperative services.
\newblock {\em SIGOPS Operating Systems Review 39}, 5 (2005).

\bibitem{amir05spread}
{\sc Amir, Y., Nita-Rotaru, C., Stanton, S., and Tsudik, G.}
\newblock Secure spread: An integrated architecture for secure group
  communication.
\newblock {\em IEEE Transactions on dependable and secure computing 2}, 3
  (2005), 248--261.

\bibitem{an2015wide}
{\sc An, K., Gokhale, A., Tambe, S., and Kuroda, T.}
\newblock Wide area network-scale discovery and data dissemination in
  data-centric publish/subscribe systems.
\newblock In {\em Middleware\/} (2015).

\bibitem{and09fawn}
{\sc Andersen, D.~G., Franklin, J., Kaminsky, M., Phanishayee, A., Tan, L., and
  Vasudevan, V.}
\newblock {FAWN: A fast array of wimpy nodes}.
\newblock In {\em SOSP\/} (2009).

\bibitem{hyperledger}
{\sc Androulaki, E., Barger, A., Bortnikov, V., Cachin, C., Christidis, K.,
  Caro, A.~D., Enyeart, D., Ferris, C., Laventman, G., Manevich, Y.,
  Muralidharan, S., Murthy, C., Nguyen, B., Sethi, M., Singh, G., Smith, K.,
  Sorniotti, A., Stathakopoulou, C., Vukolic, M., Cocco, S.~W., and Yellick,
  J.}
\newblock Hyperledger fabric: a distributed operating system for permissioned
  blockchains.
\newblock In {\em Proceedings of the Thirteenth EuroSys Conference, EuroSys
  2018, Porto, Portugal, April 23-26, 2018\/} (2018), pp.~30:1--30:15.

\bibitem{aublin15next700}
{\sc Aublin, P.-L., Guerraoui, R., Kne\v{z}evi\'{c}, N., Qu{\'e}ma, V., and
  Vukoli\'{c}, M.}
\newblock {The Next 700 BFT Protocols}.
\newblock {\em ACM Trans. Comput. Syst. 32}, 4 (2015).

\bibitem{ba11megastore}
{\sc Baker, J., Bond, C., Corbett, J.~C., Furman, J., Khorlin, A., Larson, J.,
  Leon, J.-M., Li, Y., Lloyd, A., and Yushprakh, V.}
\newblock Megastore: Providing scalable, highly available storage for
  interactive services.
\newblock In {\em CIDR\/} (2011).

\bibitem{bec11leader}
{\sc Becker, D., Junqueira, F., and Serafini, M.}
\newblock Leader election for replicated services using application scores.
\newblock In {\em Middleware\/} (2011).

\bibitem{bessani2014state}
{\sc Bessani, A., Sousa, J., and Alchieri, E.~E.}
\newblock {State machine replication for the masses with BFT-SMaRt}.
\newblock In {\em IEEE/IFIP DSN\/} (2014).

\bibitem{bezerra2016strong}
{\sc Bezerra, C.~E., and Pedone, F.}
\newblock {Strong Consistency at Scale}.
\newblock {\em IEEE Data Engineering Bulletin\/} (2016).

\bibitem{bi12spaxos}
{\sc Biely, M., Milosevic, Z., Santos, N., and Schiper, A.}
\newblock {S-Paxos: Offloading the Leader for High Throughput State Machine
  Replication}.
\newblock In {\em SRDS\/} (2012).

\bibitem{bre12cap}
{\sc Brewer, E.}
\newblock Cap twelve years later: How the "rules" have changed.
\newblock {\em Computer 45}, 2 (2012).

\bibitem{castro2002}
{\sc Castro, M., and Liskov, B.}
\newblock {Practical byzantine fault tolerance and proactive recovery}.
\newblock {\em ACM Transactions on Computer Systems 20}, 4 (2002).

\bibitem{cha07pml}
{\sc Chandra, T.~D., Griesemer, R., and Redstone, J.}
\newblock Paxos made live: An engineering perspective.
\newblock In {\em PODC\/} (2007).

\bibitem{chun07attested}
{\sc Chun, B.-G., Maniatis, P., Shenker, S., and Kubiatowicz, J.}
\newblock {Attested append-only memory}.
\newblock In {\em SOSP\/} (2007).

\bibitem{clement09making}
{\sc Clement, A., Wong, E.~L., Alvisi, L., Dahlin, M., and Marchetti, M.}
\newblock {Making Byzantine Fault Tolerant Systems Tolerate Byzantine Faults}.
\newblock In {\em NSDI\/} (2009).

\bibitem{co13spanner}
{\sc Corbett, J.~C., Dean, J., Epstein, M., Fikes, A., Frost, C., Furman,
  J.~J., Ghemawat, S., Gubarev, A., Heiser, C., Hochschild, P., Hsieh, W.,
  Kanthak, S., Kogan, E., Li, H., Lloyd, A., Melnik, S., Mwaura, D., Nagle, D.,
  Quinlan, S., Rao, R., Rolig, L., Saito, Y., Szymaniak, M., Taylor, C., Wang,
  R., and Woodford, D.}
\newblock Spanner: Google's globally distributed database.
\newblock {\em ACM Transactions on Computer Systems (TOCS) 31}, 3 (2013).

\bibitem{cowling2006hq}
{\sc Cowling, J., Myers, D., Liskov, B., Rodrigues, R., and Shrira, L.}
\newblock {HQ replication: A hybrid quorum protocol for Byzantine fault
  tolerance}.
\newblock In {\em OSDI\/} (2006).

\bibitem{Croman16}
{\sc Croman, K., Decker, C., Eyal, I., Gencer, A.~E., Juels, A., Kosba, A.~E.,
  Miller, A., Saxena, P., Shi, E., Sirer, E.~G., Song, D., and Wattenhofer, R.}
\newblock On scaling decentralized blockchains - {(A} position paper).
\newblock In {\em Financial Cryptography and Data Security - {FC} 2016
  International Workshops, BITCOIN, VOTING, and WAHC, Christ Church, Barbados,
  February 26, 2016, Revised Selected Papers\/} (2016), pp.~106--125.

\bibitem{du18beat}
{\sc Duan, S., Reiter, M.~K., and Zhang, H.}
\newblock {BEAT: Asynchronous BFT Made Practical}.
\newblock In {\em CCS\/} (2018).

\bibitem{dwork88dls}
{\sc Dwork, C., Lynch, N.~A., and Stockmeyer, L.}
\newblock Consensus in the presence of partial synchrony.
\newblock {\em JACM 35}, 2 (Apr. 1988), 288--323.

\bibitem{eyal16bitcoinng}
{\sc Eyal, I., Gencer, A.~E., Sirer, E.~G., and Van~Renesse, R.}
\newblock {Bitcoin-NG: A Scalable Blockchain Protocol}.
\newblock In {\em NSDI\/} (2016).

\bibitem{fid88times}
{\sc Fidge, C.~J.}
\newblock Timestamps in message-passing systems that preserve the partial
  ordering.
\newblock In {\em Proc. of the 11th Australian Computer Science Conference
  (ACSC'88)\/} (1988).

\bibitem{fournierzkacross}
{\sc Fournier, C.}
\newblock Running zookeeper across regions, 2012.
\newblock
  \url{http://www.elidedbranches.com/2012/12/building-global-highly-available.html}.

\bibitem{gh03gfs}
{\sc Ghemawat, S., Gobioff, H., and Leung, S.-T.}
\newblock The {G}oogle {F}ile {S}ystem.
\newblock In {\em SOSP\/} (2003).

\bibitem{gilad2017algorand}
{\sc Gilad, Y., Hemo, R., Micali, S., Vlachos, G., and Zeldovich, N.}
\newblock {Algorand: Scaling byzantine agreements for cryptocurrencies}.
\newblock In {\em SOSP\/} (2017).

\bibitem{glenden11scatter}
{\sc Glendenning, L., Beschastnikh, I., Krishnamurthy, A., and Anderson, T.}
\newblock {Scalable Consistency in Scatter}.
\newblock In {\em SOSP\/} (2011).

\bibitem{gueta2018sbft}
{\sc {Golan Gueta}, G., {Abraham}, I., {Grossman}, S., {Malkhi}, D., {Pinkas},
  B., {Reiter}, M.~K., {Seredinschi}, D.-A., {Tamir}, O., and {Tomescu}, A.}
\newblock {SBFT: a Scalable and Decentralized Trust Infrastructure}.
\newblock In {\em IEEE/IFIP DSN\/} (2019).
\newblock \url{https://arxiv.org/pdf/1804.01626.pdf}.

\bibitem{Kermarrec2000}
{\sc Guerraoui, R., Kermarrec, A.-M., Pavlovic, M., and Seredinschi, D.-A.}
\newblock {Atum : Scalable Group Communication Using Volatile Groups}.
\newblock In {\em Middleware\/} (2016).

\bibitem{guerra19cnc}
{\sc Guerraoui, R., Kuznetsov, P., Monti, M., Pavlovi\v{c}, M., and
  Seredinschi, D.-A.}
\newblock The consensus number of a cryptocurrency.
\newblock In {\em Proceedings of the 2019 ACM Symposium on Principles of
  Distributed Computing\/} (New York, NY, USA, 2019), PODC '19, ACM,
  pp.~307--316.

\bibitem{guerraoui10ring}
{\sc Guerraoui, R., Levy, R.~R., Pochon, B., and Qu{\'{e}}ma, V.}
\newblock {Throughput Optimal Total Order Broadcast for Cluster Environments}.
\newblock {\em ACM Transactions on Computer Systems (TOCS) 28}, 2 (2010).

\bibitem{guerraoui16icg}
{\sc Guerraoui, R., Pavlovic, M., and Seredinschi, D.-A.}
\newblock {Incremental Consistency Guarantees for Replicated Objects}.
\newblock In {\em OSDI\/} (2016).

\bibitem{gup16nonconsensus}
{\sc Gupta, S.}
\newblock {A Non-Consensus Based Decentralized Financial Transaction Processing
  Model with Support for Efficient Auditing}.
\newblock Master's thesis, Arizona State University, USA, 2016.

\bibitem{HerlihyW90}
{\sc Herlihy, M., and Wing, J.~M.}
\newblock Linearizability: {A} correctness condition for concurrent objects.
\newblock {\em {ACM} Trans. Program. Lang. Syst. 12}, 3 (1990), 463--492.

\bibitem{hu10zookeeper}
{\sc Hunt, P., Konar, M., Junqueira, F.~P., and Reed, B.}
\newblock Zookeeper: Wait-free coordination for internet-scale systems.
\newblock In {\em USENIX ATC\/} (2010).

\bibitem{istva16consbox}
{\sc Istv{\'a}n, Z., Sidler, D., Alonso, G., and Vukolic, M.}
\newblock Consensus in a box: Inexpensive coordination in hardware.
\newblock In {\em NSDI\/} (2016).

\bibitem{jalil14practical}
{\sc Jalili~Marandi, P., Benz, S., Pedone, F., and Birman, K.}
\newblock Practical experience report: The performance of paxos in the cloud.
\newblock {\em arXiv preprint arXiv:1404.6719\/} (2014).

\bibitem{jalili17ring}
{\sc Jalili~Marandi, P., Primi, M., Schiper, N., and Pedone, F.}
\newblock Ring paxos: High-throughput atomic broadcast.
\newblock {\em The Computer Journal 60}, 6 (2017), 866--882.

\bibitem{junq11zab}
{\sc Junqueira, F.~P., Reed, B.~C., and Serafini, M.}
\newblock {ZAB}: High-performance broadcast for primary-backup systems.
\newblock In {\em IEEE/IFIP DSN\/} (2011).

\bibitem{kapri12eve}
{\sc Kapritsos, M., Wang, Y., Quema, V., Clement, A., Alvisi, L., and Dahlin,
  M.}
\newblock {All about Eve: execute-verify replication for multi-core servers}.
\newblock In {\em OSDI\/} (2012).

\bibitem{knezevic12high}
{\sc Knezevic, N.}
\newblock {\em A High-Throughput Byzantine Fault-Tolerant Protocol}.
\newblock PhD thesis, IC/EPFL, 2012.

\bibitem{kogi16byzcoin}
{\sc Kogias, E.~K., Jovanovic, P., Gailly, N., Khoffi, I., Gasser, L., and
  Ford, B.}
\newblock Enhancing bitcoin security and performance with strong consistency
  via collective signing.
\newblock In {\em USENIX Security\/} (2016).

\bibitem{koko17omniledger}
{\sc Kokoris-Kogias, E., Jovanovic, P., Gasser, L., Gailly, N., and Ford, B.}
\newblock {OmniLedger: A Secure, Scale-Out, Decentralized Ledger}.
\newblock {\em IACR Cryptology ePrint Archive 2017\/} (2017).

\bibitem{kot07zyz}
{\sc Kotla, R., Alvisi, L., Dahlin, M., Clement, A., and Wong, E.}
\newblock Zyzzyva: {S}peculative {B}yzantine {F}ault {T}olerance.
\newblock {\em SIGOPS Operating Systems Review 41}, 6 (2007).

\bibitem{lam98paxos}
{\sc Lamport, L.}
\newblock The part-time parliament.
\newblock {\em ACM TOCS 16}, 2 (1998).

\bibitem{lamp03lowerbounds}
{\sc Lamport, L.}
\newblock Lower bounds for asynchronous consensus.
\newblock In {\em Future Directions in Distributed Computing\/} (2003).

\bibitem{lamp82byzantine}
{\sc Lamport, L., Shostak, R., and Pease, M.}
\newblock The byzantine generals problem.
\newblock {\em TOPLAS 4}, 3 (1982), 382--401.

\bibitem{liu15xft}
{\sc Liu, S., Viotti, P., Cachin, C., Qu{\'{e}}ma, V., and Vukoli{\'{c}}, M.}
\newblock {XFT: Practical Fault Tolerance Beyond Crashes}.
\newblock In {\em OSDI\/} (2016).

\bibitem{maccormick2004boxwood}
{\sc MacCormick, J., Murphy, N., Najork, M., Thekkath, C.~A., and Zhou, L.}
\newblock Boxwood: Abstractions as the foundation for storage infrastructure.
\newblock In {\em OSDI\/} (2004).

\bibitem{mar06fab}
{\sc Martin, J.~P., and Alvisi, L.}
\newblock {Fast Byzantine Consensus}.
\newblock In {\em IEEE TDSC\/} (2006).

\bibitem{matt88virtual}
{\sc Mattern, F.}
\newblock Virtual time and global states of distributed systems.
\newblock In {\em Proc. Workshop on Parallel and Distributed Algorithms\/}
  (1988).

\bibitem{miller2016honey}
{\sc Miller, A., Xia, Y., Croman, K., Shi, E., and Song, D.}
\newblock {The honey badger of BFT protocols}.
\newblock In {\em CCS\/} (2016).

\bibitem{mishra1993consul}
{\sc Mishra, S., Peterson, L.~L., and Schlichting, R.~D.}
\newblock Consul: A communication substrate for fault-tolerant distributed
  programs.
\newblock {\em Distributed Systems Engineering 1}, 2 (1993), 87.

\bibitem{mora13epaxos}
{\sc Moraru, I., Andersen, D.~G., and Kaminsky, M.}
\newblock There is more consensus in egalitarian parliaments.
\newblock In {\em SOSP\/} (2013).

\bibitem{ongaro2014consensus}
{\sc Ongaro, D.}
\newblock {\em Consensus: Bridging theory and practice}.
\newblock PhD thesis, Stanford University, 2014.

\bibitem{ongaro14search}
{\sc Ongaro, D., and Ousterhout, J.}
\newblock In search of an understandable consensus algorithm.
\newblock In {\em USENIX ATC\/} (2014).

\bibitem{bftshome}
{\sc Online}.
\newblock {Bft-SMaRt home page}.
\newblock \url{http://bft-smart.github.io/library/}.

\bibitem{bftspkg}
{\sc Online}.
\newblock bft-smart/library, commit 3af266d4 dated {J}ul 12, 2016.
\newblock \url{https://github.com/bft-smart/library/commit/3af266d4}.

\bibitem{etcd}
{\sc Online}.
\newblock {CoreOS}/etcd.
\newblock \url{https://coreos.com/etcd/}.

\bibitem{etcduse}
{\sc Online}.
\newblock etcd production users.
\newblock
  \url{https://github.com/coreos/etcd/blob/master/Documentation/production-users.md}.

\bibitem{etcdbatching}
{\sc Online}.
\newblock etcd/etcdserver/raft.go.
\newblock
  \url{https://github.com/coreos/etcd/blob/master/etcdserver/raft.go\#L48}.

\bibitem{softlayerDCs}
{\sc Online}.
\newblock Softlayer.
\newblock \url{http://www.softlayer.com/our-platform}.

\bibitem{upstart}
{\sc Online}.
\newblock {Upstart: event-based init deamon}.
\newblock \url{http://upstart.ubuntu.com}.

\bibitem{zkpkg}
{\sc Online}.
\newblock Zoo{K}eeper ({U}buntu package), high-performance coordination service
  for distributed applications.
\newblock \url{http://packages.ubuntu.com/trusty/zookeeper}.

\bibitem{kafkaHL}
{\sc Online}.
\newblock {Hyperledger Architecture, Volume 1}, August 2017.
\newblock
  \url{https://www.hyperledger.org/wp-content/uploads/2017/08/HyperLedger_Arch_WG_Paper_1_Consensus.pdf}.

\bibitem{poke15dare}
{\sc Poke, M., and Hoefler, T.}
\newblock {DARE: High-Performance State Machine Replication on RDMA Networks}.
\newblock In {\em HPDC\/} (2015).

\bibitem{Visigoth}
{\sc Porto, D., Leit{\~{a}}o, J., Li, C., Clement, A., Kate, A., Junqueira,
  F.~P., and Rodrigues, R.}
\newblock Visigoth fault tolerance.
\newblock In {\em EuroSys\/} (2015).

\bibitem{qi13espresso}
{\sc Qiao, L., Auradar, A., Beaver, C., Brandt, G., Gandhi, M., Gopalakrishna,
  K., Ip, W., Jgadish, S., Lu, S., Pachev, A., Ramesh, A., Surlaker, K.,
  Sebastian, A., Shanbhag, R., Subramaniam, S., Sun, Y., Topiwala, S., Tran,
  C., Westerman, J., Zhang, D., Das, S., Quiggle, T., Schulman, B., Ghosh, B.,
  Curtis, A., Seeliger, O., and Zhang, Z.}
\newblock {On Brewing Fresh Espresso: LinkedIn’s Distributed Data Serving
  Platform}.
\newblock In {\em SIGMOD\/} (2013).

\bibitem{santos2012tuning}
{\sc Santos, N., and Schiper, A.}
\newblock Tuning paxos for high-throughput with batching and pipelining.
\newblock In {\em International Conference on Distributed Computing and
  Networking\/} (2012), Springer, pp.~153--167.

\bibitem{sh11crdt}
{\sc Shapiro, M., Pregui{\c{c}}a, N., Baquero, C., and Zawirski, M.}
\newblock Conflict-free replicated data types.
\newblock In {\em Stabilization, Safety, and Security of Distributed Systems}.
  Springer, 2011.

\bibitem{shraer12dynamic}
{\sc Shraer, A., Reed, B., Malkhi, D., and Junqueira, F.~P.}
\newblock Dynamic {R}econfiguration of {P}rimary/{B}ackup {C}lusters.
\newblock In {\em USENIX ATC\/} (2012).

\bibitem{sousa15wheat}
{\sc Sousa, J., and Bessani, A.}
\newblock {Separating the WHEAT from the Chaff: An Empirical Design for
  Geo-Replicated State Machines}.
\newblock In {\em SRDS\/} (2015).

\bibitem{sou18byzantine}
{\sc Sousa, J., Bessani, A., and Vukolic, M.}
\newblock {A Byzantine Fault-Tolerant Ordering Service for the Hyperledger
  Fabric Blockchain Platform}.
\newblock In {\em DSN\/} (2018).

\bibitem{tange2011gnu}
{\sc Tange, O., et~al.}
\newblock {GNU} parallel - {The Command-Line Power Tool}.
\newblock {\em The USENIX Magazine 36}, 1 (2011), 42--47.

\bibitem{rocket18avalanche}
{\sc Team-Rocket}.
\newblock {Snowflake to Avalanche: A Novel Metastable Consensus Protocol Family
  for Cryptocurrencies}.
\newblock {\em White Paper\/} (2018).
\newblock Revision: 05/16/2018 21:51:26 UTC.

\bibitem{van12byzantine}
{\sc Van~Renesse, R., Ho, C., and Schiper, N.}
\newblock Byzantine chain replication.
\newblock In {\em OPODIS\/} (2012).

\bibitem{van15vive}
{\sc Van~Renesse, R., Schiper, N., and Schneider, F.~B.}
\newblock Vive la diff{\'e}rence: {P}axos vs. {V}iewstamped {R}eplication vs.
  {ZAB}.
\newblock {\em IEEE Transactions on Dependable and Secure Computing 12}, 4
  (2015), 472--484.

\bibitem{van04chain}
{\sc Van~Renesse, R., and Schneider, F.~B.}
\newblock Chain replication for supporting high throughput and availability.
\newblock In {\em OSDI\/} (2004).

\bibitem{Vogels09}
{\sc Vogels, W.}
\newblock Eventually consistent.
\newblock {\em Commun. {ACM} 52}, 1 (2009), 40--44.

\bibitem{vuko15quest}
{\sc Vukoli{\'c}, M.}
\newblock The {Q}uest for {S}calable {B}lockchain {F}abric: {P}roof-of-work vs.
  {BFT} {R}eplication.
\newblock In {\em International Workshop on Open Problems in Network
  Security\/} (2015), Springer, pp.~112--125.

\bibitem{weste09latency}
{\sc Wester, B., Cowling, J., Nightingale, E.~B., Chen, P.~M., Flinn, J., and
  Liskov, B.}
\newblock {Tolerating latency in replicated state machines through client
  speculation}.
\newblock In {\em NSDI\/} (2009).

\bibitem{wu2013spanstore}
{\sc Wu, Z., Butkiewicz, M., Perkins, D., Katz-Bassett, E., and Madhyastha,
  H.~V.}
\newblock Spanstore: Cost-effective geo-replicated storage spanning multiple
  cloud services.
\newblock In {\em SOSP\/} (2013).

\end{thebibliography}

\end{document}